\documentclass[numbers]{sigplanconfzyw}

\usepackage[ruled]{algorithm2e}

\usepackage[pdftex]{graphicx}
\DeclareGraphicsExtensions{.pdf,.jpeg,.png}

\usepackage{booktabs}
\usepackage{color}
\usepackage[colorlinks=true,citecolor=blue,linkcolor=blue,urlcolor=blue]{hyperref}

\usepackage[mathscr]{euscript}

\usepackage{amsmath}
\usepackage{amssymb}
\usepackage{latexsym}
\usepackage{proof}

\usepackage{isabelle,isabellesym}

\newcommand{\isacodeftsz}{\footnotesize}  

\isabellestyle{it}

\newtheorem{theorem}{Theorem}
\newtheorem{lemma}{Lemma}

\newtheorem{definition}{Definition}

\newcommand{\sectprefix}{{Section}}
\newcommand{\subsectprefix}{{Subsection}}
\newcommand{\figprefix}{{Fig.}}

\newcommand{\defprefix}{Definition}
\newcommand{\lemmaprefix}{Lemma}
\newcommand{\theoremprefix}{Theorem}

\newcommand{\problem}[1]{}
\newcommand{\todo}[1]{}

\newcommand{\llbrace}{\lbrace\mkern-4.5mu\mid}
\newcommand{\rrbrace}{\mid\mkern-4.5mu\rbrace}

\newcommand{\slang}{$\pi$-Core} 

\newcommand{\cmdfont}[1]{\textbf{#1}} 

\newcommand{\defi}{\triangleq}

\newcommand{\cmdskip}{\cmdfont{Skip}}
\newcommand{\cmdbasic}[1]{\cmdfont{Basic} \ {#1}}
\newcommand{\cmdseq}[2]{\cmdfont{Seq} \ {#1} \ {#2}}
\newcommand{\cmdcond}[3]{\cmdfont{Cond} \ {#1} \ {#2} \ {#3}}
\newcommand{\cmdwhile}[2]{\cmdfont{While} \ {#1} \ {#2}}
\newcommand{\cmdawait}[2]{\cmdfont{Await} \ {#1} \ {#2}}
\newcommand{\cmdnondt}[1]{\cmdfont{Nondt} \ {#1}}
\newcommand{\cmdnone}{\cmdfont{None}}

\newcommand{\event}[1]{\cmdfont{BasicEvt} \ {#1}}
\newcommand{\anonevt}[1]{\textbf{AnonyEvt} \ {#1}}

\newcommand{\evtcomp}{\oplus}
\newcommand{\evtsys}[3]{{#1} \ \evtcomp \ {#2} \ ... \ \evtcomp \ {#3}}
\newcommand{\evtseq}[2]{{#1};{#2}}

\newcommand{\parsysc}{\symbCore \rightarrow \symbevtsys}

\newcommand{\symbprog}{P}
\newcommand{\symbbexp}{b}
\newcommand{\symbEvt}{\mathcal{E}}
\newcommand{\symbevt}{ev}
\newcommand{\symbevtbd}{\alpha}
\newcommand{\symbevtsys}{\mathcal{S}}
\newcommand{\symbpes}{\mathcal{PS}}
\newcommand{\symbState}{S}
\newcommand{\symbstate}{s}

\newcommand{\symbevtctx}{x}
\newcommand{\symbConf}{\Delta}
\newcommand{\symbconf}{\mathcal{C}}
\newcommand{\symbpcomp}{c}

\newcommand{\symbact}{t}
\newcommand{\symbCore}{\mathcal{K}}
\newcommand{\symbcore}{\kappa}
\newcommand{\symbactk}{\delta}

\newcommand{\actk}[2]{{#1}@{#2}}

\newcommand{\symbDomain}{\mathcal{D}}
\newcommand{\symbdomain}{d}
\newcommand{\symbdoms}{D}
\newcommand{\symbSM}{\mathcal{M}}
\newcommand{\symbAction}{A}
\newcommand{\symbaction}{a}
\newcommand{\symbactions}{as}
\newcommand{\symbspec}{\sharp}

\newcommand{\tran}[1]{\stackrel{{#1}}{\longrightarrow}}
\newcommand{\trank}[2]{\stackrel{\actk{#1}{#2}}{\longrightarrow}}

\newcommand{\dsim}[1]{\stackrel{{#1}}{\sim}}

\newcommand{\interf}{\leadsto}

\newcommand{\reachablef}{\mathcal{R}}

\newcommand{\execution}[2]{{#1} \triangleright {#2}}
\newcommand{\equidom}[3]{{#1}\stackrel{#2}{\sim}{#3}}
\newcommand{\equidoms}[3]{{#1}\stackrel{#2}{\approx}{#3}}
\newcommand{\ssequidom}[3]{{#1}\stackrel{#2}{\bumpeq}{#3}}
\newcommand{\equivexec}[5]{\ssequidom{\execution{#1}{#2}}{#3}{\execution{#4}{#5}}}

\newcommand{\evtran}{\stackrel{e}{\longrightarrow}}
\newcommand{\symbtran}{\mathbf{t}} 
\newcommand{\compfun}{\Psi}
\newcommand{\symbcomp}{\varpi}
\newcommand{\symbcompk}{\widehat{\symbcomp}}
\newcommand{\compsim}[2]{{#1} \asymp {#2}}
\newcommand{\compconjoin}[2]{{#1} \propto {#2}}
\newcommand{\serialize}{\lll}
\newcommand{\Serialize}[2]{{#1} \serialize {#2}}
\newcommand{{\compstps}}{\Psi_\symbpes}
\newcommand{{\compstes}}{\Psi_\symbevtsys}
\newcommand{{\compste}}{\Psi_\symbEvt}
\newcommand{{\compstp}}{\Psi_\symbprog}
\newcommand{\assumefun}{A}
\newcommand{\commitfun}{C}
\newcommand{\rgcond}[4]{\langle #1, #2, #3, #4 \rangle}
\newcommand{\rgconddefault}{\rgcond{pre}{R}{G}{pst}}
\newcommand{\RGSAT}[2]{\models \ {#1} \ \mathbf{sat} \ {#2}}
\newcommand{\rgsat}[2]{\vdash \ {#1} \ \mathbf{sat} \ {#2}}

\makeatletter
\newcommand{\superimpose}[2]{%
  {\ooalign{$#1\@firstoftwo#2$\cr\hfil$#1\@secondoftwo#2$\hfil\cr}}}
\makeatother
\newcommand{\ninterf}{\mathrel{\mathpalette\superimpose{{\slash}{\leadsto}}}}

\begin{document}

\setlength{\pdfpageheight}{\paperheight}
\setlength{\pdfpagewidth}{\paperwidth}

\conferenceinfo{POPL 2016 Submission}{July, 2016}
\copyrightyear{2016}
\copyrightdata{978-1-nnnn-nnnn-n/yy/mm}
\copyrightdoi{nnnnnnn.nnnnnnn}



\title{Event-based Compositional Reasoning of Information-Flow Security for Concurrent Systems}

\authorinfo{Yongwang Zhao}
           {Nanyang Technological University, Singapore \\ Beihang University, China}
           {ywzhao@ntu.edu.sg, zhaoyw@buaa.edu.cn}

\authorinfo{David San\'an \quad Fuyuan Zhang \quad  Yang Liu}
           {Nanyang Technological University, Singapore}
           {\{fuzh,sanan,yangliu\}@ntu.edu.sg}

\maketitle

\begin{abstract}
High assurance of information-flow security (IFS) for concurrent systems is challenging. A promising way for formal verification of concurrent systems is the rely-guarantee method. However, existing compositional reasoning approaches for IFS concentrate on language-based IFS. It is often not applicable for system-level security, such as multicore operating system kernels, in which secrecy of actions should also be considered. On the other hand, existing studies on the rely-guarantee method are basically built on concurrent programming languages, by which semantics of concurrent systems cannot be completely captured in a straightforward way. 

In order to formally verify state-action based IFS for concurrent systems, we propose a rely-guarantee-based compositional reasoning approach for IFS in this paper. 
We first design a language by incorporating ``Event'' into concurrent languages and give the IFS semantics of the language. As a primitive element, events offer an extremely neat framework for modeling system and are not necessarily atomic in our language. For compositional reasoning of IFS, we use rely-guarantee specification to define new forms of unwinding conditions (UCs) on events, i.e., event UCs. By a rely-guarantee proof system of the language and the soundness of event UCs, we have that event UCs imply IFS of concurrent systems. In such a way, we relax the atomicity constraint of actions in traditional UCs and provide a compositional reasoning way for IFS in which security proof of systems can be discharged by independent security proof on individual events. 
Finally, we mechanize the approach in Isabelle/HOL and develop a formal specification and its IFS proof for multicore separation kernels as a study case according to an industrial standard -- ARINC 653. 
\end{abstract}




\section{Introduction}

Information-flow security (IFS) \cite{sabel03} deals with the problem of preventing improper release and modification of information in complex systems.
It has been studied at multiple levels of abstraction, such as the application level, the operating system level, and the hardware level. 
Nowadays critical and high-assurance systems are designed for multi-core architectures where multiple subsystems are running in parallel. 
For instance, recent microkernels like XtratuM \cite{Carr14} are shared-variable concurrent systems, where the scheduler and system services may be executed simultaneously on different cores of a processor. 
Information-flow security of concurrent systems is an increasingly important and challenging problem. 

Traditionally, language-based IFS \cite{sabel03} at the application level defines security policies of computer programs and concerns the data confidentiality to prevent information leakage from \emph{High} variables to \emph{Low} ones. However, language-based IFS is often not applicable for system-level security, because (1) in many cases it is impossible to classify \emph{High} and \emph{Low} variables; (2) data confidentiality is a weak property and is not enough for system-level security; and (3) language-based IFS is not able to deal with intransitive policies straightforwardly. Therefore, state-action based IFS \cite{rushby92,Oheimb04}, which can deal with data confidentiality and secrecy of actions together, is usually adopted in formal verification of microkernels \cite{Klein14}, separation kernels \cite{verbeek15,dam13,Zhao16,richards10}, and microprocessors \cite{wilding10}. 

The state-action based IFS is defined on a state machine and security proof is discharged by proving a set of unwinding conditions (UCs) \cite{rushby92} that examine individual transitions of the state machine. Although compositional reasoning of language-based IFS has been studied \cite{Mantel11,Murray16}, the lack of compositional reasoning of state-action based IFS prevents applying this approach to formally verifying large and concurrent systems. The rely-guarantee method \cite{Jones83,Xu97} represents a fundamental compositional method for correctness proofs of concurrent systems with shared variables. However, the existing studies on the rely-guarantee method concentrate on concurrent programs (e.g. \cite{Xu97, Nieto03, LiangFF12}) which are basically represented in imperative languages with extensions of concurrency. Concurrent systems are not just concurrent programs, for example, the occurrence of exceptions/interrupts from hardware is beyond the scope of programs. The existing languages and their rely-guarantee proof systems do not provide a straightforward way to specify and reason concurrent systems. Moreover, the formalization of concurrent programs in existing rely-guarantee methods is at source code level. Choosing the right level of abstraction instead of the low-level programs allows both precise information flow analysis and high-level programmability. 

Finally, IFS and its formal verification on multicore separation kernels are challenging. As an important sort of concurrent systems, multicore separation kernels establish an execution environment, which enables different criticality levels to share a common set of physical resources, by providing to their hosted applications spatial/temporal separation and controlled information flow. 
The security of separation kernels is usually achieved by the Common Criteria (CC) \cite{CC} evaluation, in which formal verification of IFS is mandated for high assurance levels.
Although formal verification of IFS on monocore microkernels and separation kernels has been widely studied (e.g. \cite{Murray12, Murray13, Heitmeyer08, verbeek15, dam13, Zhao16, richards10}), to the best of our knowledge, there is no related work about compositional reasoning of IFS on multicore operating systems in the literature.

To address the above problems, we propose a rely-guarantee-based compositional reasoning approach for verifying information-flow security of concurrent systems in this paper. 
We first propose an event-based concurrent language -- {\slang}, which combines elements of concurrent programming languages and system specification languages. In {\slang}, an event system represents a single-processing system and is defined by a set of \emph{events}, each of which defines the state transition that can occur under certain circumstances. A concurrent system is defined as a parallel event system on shared states, which is the parallel composition of event systems. Due to the shared states and concurrent execution of event systems, the execution of events in a parallel event system is in an interleaved manner. Then, we define the IFS semantics of {\slang} which includes IFS properties and an unwinding theorem to show that UCs examining small-step and atomic actions imply the IFS. 
In order to compositionally verify IFS of {\slang}, we provide a rely-guarantee proof system for {\slang} and prove its soundness. Next, we use rely-guarantee specification to define new forms of UCs on events, i.e., event UCs, which examines big-step and non-atomic events. A soundness theorem for event UCs shows that event UCs imply the small-step UCs, and thus the IFS. In such a way, we provide a compositional reasoning for IFS in which security proof of systems can be discharged by local security proof on events. 
In detail, we make the following contributions:
\begin{itemize}
\item We propose an event-based language {\slang} and its operational semantics by incorporating ``Event'' into concurrent programming languages. The language could be used to create formal specification of concurrent systems as well as to design and implement the system. Beside the semantics of software parts, the behavior of hardware parts of systems could be specified. 

\item We define the IFS semantics of {\slang} on a state machine, which is transformed from {\slang}. A transition of the state machine represents an atomic execution step of a parallel event system. A set of IFS properties and small-step UCs are defined on the state machine. We prove an unwinding theorem, i.e., small-step UCs imply the IFS of concurrent systems. 

\item We build a rely-guarantee proof system for {\slang} and prove its soundness. This work is the first effort to study the rely-guarantee method for system-level concurrency in the literature. We provide proof rules for both parallel composition of event systems and nondeterministic occurrence of events.  
Although, we use the proof system for compositional reasoning of IFS in this paper, it is possible to use the proof system for the functional correctness and safety of concurrent systems. 

\item We propose a rely-guarantee-based approach to compositionally verifying IFS of {\slang}. Based on the rely-guarantee specification of events, we define new forms of UCs on big-step and non-atomic events. We prove the soundness, i.e., event UCs imply the small-step UCs of {\slang}, and thus the security. This work is the first effort to study compositional reasoning of state-action based IFS. 

\item We formalize the {\slang} language, the IFS semantics, the rely-guarantee proof system, and compositional reasoning of IFS in the Isabelle/HOL theorem prover \footnote{The sources files in Isabelle are available as supplementary material. The official web address will be available in camera ready version.}. All results have been proved in Isabelle/HOL. We also create a concrete syntax for {\slang} which is convenient to specify and verify concurrent systems. 

\item By the compositional approach and its implementation in Isabelle/HOL, we develop a formal specification and its IFS proof of multicore separation kernels according to the ARINC 653 standard. This work is the first effort to formally verify the IFS of multicore separation kernels in the literature. 
\end{itemize}

In the rest of this paper, we first give an informal overview in {\sectprefix} \ref{sect:inf_ov} which includes the background, problems and challenges in this work, and an overview of our approach. Then we define the {\slang} language in {\sectprefix} \ref{sect:pclang} and its IFS semantics in {\sectprefix} \ref{sect:ifs}. The rely-guarantee proof system is presented in {\sectprefix} \ref{sect:rgproof}. In {\sectprefix} \ref{sect:rgifs}, we discuss the rely-guarantee approach of IFS. The study case of multicore separation kernels is presented in {\sectprefix} \ref{sect:case}. Finally we discuss related work and conclude in {\sectprefix} \ref{sect:rw_concl}. 

\section{Informal Overview}
\label{sect:inf_ov}
In this section, we first present technical background, problems and challenges in this work. Then, we overview our approach. 

\subsection{Background}
\paragraph{Rely-guarantee method.}
Rely-guarantee \cite{Jones83,Xu97} is a compositional proof system that extends the specification of concurrent programs with rely and guarantee conditions. 
The two conditions are predicates over a pair of states and characterizes, respectively, how the environment interferes with the program under execution and how the program guarantees to the environment.
Therefore, the specification of a program is a quadruple $(p, R, G, q)$, where $p$ and $q$ are pre- and post-conditions, and $R$ and $G$ are rely and guarantee conditions. 
A program satisfies its specification if, given an initial state satisfying $p$ and an environment whose transitions satisfy $R$, each atomic transition made by the program satisfies $G$ and the final state satisfies $q$.
A main benefit of this method is compositionality, i.e., the verification of large concurrent programs can be reduced to the independent verification of individual subprograms.

\paragraph{Information-flow security.}
The notion \emph{noninterference} is introduced in \cite{Goguen82} in order to provide a formal foundation for the specification and analysis of IFS policies. 
The idea is that a security domain $u$ is noninterfering with a domain $v$ if no action performed by $u$ can influence the subsequent outputs seen by $v$.
Language-based IFS \cite{sabel03} defines security policies of programs and handles two-level domains: \emph{High} and \emph{Low}. The variables of programs are assigned either \emph{High} or \emph{Low} labels. Security hereby concerns the data confidentiality to prevent information leakage, i.e. variations of the \emph{High}-level data should not cause a variation of the \emph{Low}-level data. Intransitive policies \cite{rushby92} cannot be addressed by traditional language-based IFS \cite{Oheimb04}. 
This problem is solved in \cite{rushby92}, where noninterference is defined in a state-action manner. The state-action based noninterfernce concerns the visibility of \emph{actions}, i.e. the secrets that actions introduce in the system state. It is usually chosen for verifying system-level security, such as general purpose operating systems and separation kernels \cite{Murray12}.  
Language-based IFS is generalized to arbitrary multi-domain policies in \cite{Oheimb04} as a new state-action based notion \emph{nonleakage}. In \cite{Oheimb04}, nonleakage and the classical noninterference are combined as a new notion \emph{noninfluence}, which considers both the data confidentiality and the secrecy of actions. These properties have been instantiated for operating systems in \cite{Murray12} and formally verified on the seL4 monocore microkernel \cite{Murray13}. 


\subsection{Problems and Challenges}

\paragraph{Rely-guarantee languages are not straightforward for systems.}
The studies on the rely-guarantee method focus on compositional reasoning of concurrent programs. Hence, the languages used in rely-guarantee methods (e.g. \cite{Xu97, Nieto03, LiangFF12}) basically extend imperative languages by parallel composition. The semantics of a system cannot be completely captured by these programming languages. For instance, interrupt handlers (e.g., system calls and scheduling) in microkernels are programmed in C language. It is beyond the scope of C language when and how the handlers are triggered. However, it is necessary to capture this kind of system behavior for the security of microkernels. The languages in the rely-guarantee method do not provide a straightforward way to specify and verify such behavior in concurrent systems. 
\citet{Jones15} mention that employing  ``Actions'' \cite{Back91} or ``Events'' \cite{Abrial07} into rely-guarantee can offer an extremely neat framework for modelling systems. On the other hand, nondeterminism is also necessary for system specification at abstraction levels, which is also not supported by languages in the rely-guarantee method. 

\paragraph{Incorporating languages and state machines for IFS. }
The rely-guarantee method defines a concurrent programming language and a set of proof rules w.r.t. semantics of the language. The rely/guarantee condition is a set of state pairs, where the action triggering the state transition is not taken into account. It is the same as language-based IFS which defines the security based on the state trace. However, state-action based IFS is defined on a state machine and takes actions into account for secrecy of actions. Rely-guarantee-based compositional reasoning of state-action based IFS requires the connection between the programming language and the state machine. We should create the relation of program execution and rely/guarantee conditions to the actions.

\paragraph{Compositionality of state-action based IFS is unclear.}
Language-base IFS concerns information leakage among state variables and is a weaker property than state-action based IFS. Compositional verification of language-based IFS has been studied (e.g. \cite{Mantel11,Murray16}) before. As a strong security property, compositionality of state-action based IFS for concurrent system is still unclear. The standard proof of state-action based IFS is discharged by proving
a set of unwinding conditions that examine individual transitions of the system. Here, the individual transition is executed in an atomic manner. Directly applying the unwinding conditions to concurrent systems may lead to explosion of the proof space due to the interleaving. The atomicity of actions on which unwinding conditions are defined has to be relaxed for compositional reasoning such that unwinding conditions can be defined on more coarse-grained level of granularity.

\paragraph{Verifying IFS of multicore microkernels is difficulty.}
Formal verification of IFS on monocore microkernels has been widely studied (e.g. \cite{Murray12, Murray13, Heitmeyer08, verbeek15, dam13, Zhao16, richards10}). IFS of seL4 assumes that interrupts are disabled in kernel mode to avoid in-kernel concurrency \cite{Murray13}. The assumption simplifies the security proof by only examining big-step actions (e.g., system calls and scheduling). In multicore microkernels, the kernel code is concurrently executed on different processor cores with the shared memory. The verification approaches for monocore microkernels are not applicable for multicore.

\subsection{Our Approach}
In order to provide a rely-guarantee proof system for concurrent systems, we first introduce \emph{events} into programming languages in the rely-guarantee method. An example of events in the concrete syntax is shown in {\figprefix} \ref{fig:event_examp}. An event is actually a non-atomic and parametrized state transition of systems with a set of guard conditions to constrain the type and value of parameters, and current state. The body of an event defines the state transition and is represented by imperative statements. We provide a special parameter $\symbcore$ for events to indicate the execution context of an event, i.e., on which single-processing system that the event is executing. For instance, the $\symbcore$ could be used to indicate the current processor core in multicore systems. 

\begin{figure}
\begin{isabellec}
\footnotesize
\isacommand{EVENT} evt1\ ps @ $\symbcore$ \isacommand{WHERE} \isanewline
\ \ \ $ps \ typeof \ [int,int] \ \wedge $ \ \ \ /*type of parameters*/\isanewline
\ \ \ $ps!0 \geq 0 \wedge ps!1 \geq 0 \ \wedge$ \ \ /*constraints of parameters*/\isanewline
\ \ \ $size(buf) \neq 0$ \ \ \ \ \ \ \ \ \ /*constraints of state*/ \isanewline
\isacommand{THEN} \isanewline
\ \ \ \isacommand{AWAIT}(lock = 0) \isacommand{THEN} \isanewline
\ \ \ \ \ \ lock := 1 \isanewline
\ \ \ \isacommand{END};; \isanewline
\ \ \ /*do something here*/ \isanewline
\ \ \ lock := 0 \isanewline
\isacommand{END}
\end{isabellec}
\caption{An Example of Event}
\label{fig:event_examp}
\end{figure}

An event system represents the behavior of a single-processing system and has two forms of event composition, i.e. \emph{event sequence} and \emph{event set}. The event sequence models the sequential execution of events. The event set models the nondeterministic occurrence of events, i.e., events in this set can occur when the guard condition is satisfied. The parallel composition of event systems is fine-grained since small-step actions in events are interleaved in semantics of {\slang}. This relaxes the atomicity constraint of events in other approaches (e.g. Event-B \cite{Abrial07}).
It is obvious that concurrent programs represented by the languages in \cite{Xu97, Nieto03, LiangFF12} could be represented by {\slang} too.

State-action based IFS is defined and proved based on a state machine. We construct a state machine from a parallel event system in {\slang}. Each action of the machine is a small-step action of events. To relate the small step to the action, each transition rule in operational semantics of {\slang} has an action label to indicate the kind of the transition. The action label shows the information about action type and in which event system the action executes. On the other hand, we add a new element, i.e. event context, in the configuration in the semantics. The event context is a function to indicate which event is currently executing in each event system. 


Then, IFS of {\slang} is defined on the state machine. In this paper, we use two-level unwinding conditions, i.e. small-step and event unwinding conditions. The small-step UCs examine small steps in events, which is atomic. The unwinding theorem shows that satisfaction of small-step UCs implies the security. This is the IFS semantics of {\slang} by following traditional IFS. The problem of directly applying the unwinding theorem is the explosion of proof space due to interleaving and the small-step conditions. A solution is to enlarge the granularity to the event level, and thus we define the event UCs of {\slang}. Since the guarantee condition of an event characterizes how the event modifies the environment, the event UCs are defined based on the guarantee condition of events. Finally, the compositionality of state-action based IFS means that if all events defined in a concurrent system satisfy the event UCs and the system is closed, then the system is secure. We conclude this by the soundness of event UCs, i.e., event UCs imply the small-step UCs in {\slang}.

\section{The {\slang} Language}
\label{sect:pclang}
This section introduces the {\slang} language including its abstract syntax, operational semantics, and computations.

\subsection{Abstract Syntax}

By introducing ``Events'' into concurrent programming languages, we create a language with four levels of elements, i.e., \emph{programs} represented by programming languages, \emph{events} constructed based on programs, \emph{event systems} composed by events, and \emph{parallel event systems} composed by event systems. The abstract syntax of {\slang} is shown in {\figprefix} \ref{fig:syntax}. 

\begin{figure}
\footnotesize
\textbf{Program}:
\begin{equation*}
\begin{aligned}
\symbprog \ ::= & \ \cmdbasic{f} \ | \ \cmdseq{\symbprog_1}{\symbprog_2} \ | \ \cmdcond{\symbbexp}{\symbprog_1}{\symbprog_2} \\
| & \ \cmdwhile{\symbbexp}{\symbprog} \ | \ \cmdawait{\symbbexp}{\symbprog} \ | \ \cmdnondt{r} 
\end{aligned}
\end{equation*}

\textbf{Event}:
\begin{equation*}
\begin{aligned}
\symbEvt \ ::= & \ \event{\symbevtbd} & (Basic \ Event)  \\ 
 | & \ \anonevt{\symbprog} & (Anonymous \ Event)
\end{aligned}
\end{equation*}

\textbf{Event System}:
\begin{equation*}
\begin{aligned}
\symbevtsys \ ::= & \ \evtsys{\symbEvt_0}{\symbEvt_1}{\symbEvt_n} & (Event \ Set) \\
| & \ \evtseq{\symbEvt}{\symbevtsys} & (Event \ Sequence)
\end{aligned}
\end{equation*}

\textbf{Parallel Event System}:
\begin{equation*}
\begin{split}
\symbpes \ ::= \ \parsysc  
\end{split}
\end{equation*}
\caption{Abstract Syntax of the {\slang} Language}
\label{fig:syntax}
\end{figure}

The syntax of programs is intuitive and is used to describe the behavior of events. The {$\cmdbasic{f}$} command represents an atomic state transformation, for example, an assignment and the {\cmdskip} command. The {$\cmdawait{\symbbexp}{\symbprog}$} command executes program $\symbprog$ atomically whenever boolean condition $\symbbexp$ holds. The $\cmdnondt{r}$ command defines the potential next states via the state relation $r$. It can be used to model nondeterministic choice. The rest are well-known. 

An event is actually a parametrized program to represent the state change of an event system. 
In an event, $\symbevtbd$ with the type of $(p \times \symbCore) \rightarrow (g \times \symbprog)$ is an event specification, 
where $p$ is the parameters, $\symbCore$ indicates the label of an event system, $g$ is the guard condition of the event, and $\symbprog$ is a program which is the body of the event. An event $\event{\symbevtbd}$ can occur under concrete parameters $p$ in event system $\symbcore$ when its guard condition (i.e. $fst(\symbevtbd(p,\symbcore))$) is true in current state. Then, it behaves as an anonymous event $\anonevt{(snd(\symbevtbd(p,\symbcore)))}$. An anonymous event is actually a wrapper of a program to represent the intermediate specification during execution of events.

The event system indeed constitutes a kind of state transition system. It has two forms of event composition, i.e. \emph{event sequence} and \emph{event set}. For an event set, when the guard conditions of some events are true, then one of the corresponding events necessarily occurs and the state is modified accordingly. When the occurred event is finished, the guard conditions are checked again, and so on. For an event sequence $\evtseq{\symbEvt}{\symbevtsys}$, when the guard condition of event $\symbEvt$ is true, then $\symbEvt$ necessarily occurs and the state is modified accordingly, finally it behaves as event system $\symbevtsys$. 

A concurrent system is modeled by a parallel event system, which is the parallel composition of event systems. The parallel composition is a function from $\symbCore$ to event systems. 
Note that a model eventually terminates is not mandatory. As a matter of fact, most of the systems we study run forever.

We introduce an auxiliary function to query all events defined in event systems and parallel event systems as follows. 
\begin{equation*}
\begin{aligned}
&\left\{
\begin{aligned}
& evts(\evtsys{\symbEvt_0}{\symbEvt_1}{\symbEvt_n}) = \{\symbEvt_0, \symbEvt_1, ..., \symbEvt_n\} \\
& evts(\evtseq{\symbEvt}{\symbevtsys}) = \{\symbEvt\} \cup evts(\symbevtsys)
\end{aligned}
\right. \\
& \ \ \ \ evts(\symbpes) \defi \bigcup_\symbcore evts(\symbpes(\symbcore))
\end{aligned}
\end{equation*}

\subsection{Operational Semantics}
Semantics of {\slang} is defined via transition rules between configurations. A configuration $\symbconf$ is defined as a triple $(\symbspec, \symbstate, \symbevtctx)$, where $\symbspec$ is a specification (e.g., a program, an event, an event system, or a parallel event system), $\symbstate$ is a state, and $\symbevtctx : \symbCore \rightarrow \symbEvt$ is an event context. The event context indicates which event is currently executed in an event system. We use $\symbspec_\symbconf$, $\symbstate_\symbconf$, and $\symbevtctx_\symbconf$ to represent the three parts of a configuration $\symbconf$ respectively. 

A system can perform two kinds of transitions: \emph{action transitions}, performed by the system itself, and \emph{environment transitions}, performed by a different system of the parallel composition or by an arbitrary environment. 

A transition rule of actions has the form $(\symbspec_1, \symbstate_1, \symbevtctx_1) \tran{\symbactk} (\symbspec_2, \symbstate_2, \symbevtctx_2)$, where $\symbactk = \actk{\symbact}{\symbcore}$ is a label indicating the kind of transition. $\symbact ::= c \ | \ \symbEvt$, where $c$ is a program action and $\symbEvt$ is the occurrence of event $\symbEvt$. $@\symbcore$ means that the action $\symbactk$ occurs in event system $\symbcore$. 
A rule of environment transition has the form $(\symbspec, \symbstate, \symbevtctx) \evtran (\symbspec, \symbstate', \symbevtctx')$, where $e$ is the label of environment transition. Intuitively, a transition made by the environment may change the state and the event context but not the specification. 

Transition rules of actions are shown in {\figprefix} \ref{fig:semantics}. 
The transition rules of programs are mostly standard. The $\tran{\symbpcomp^*}$ in the $\textsc{Await}$ rule is the reflexive transitive closure of $\tran{\symbpcomp}$. The program action modifies the state but not the event context. The execution of $\anonevt{\symbprog}$ mimics program $\symbprog$. The $\textsc{BasicEvt}$ rule shows the occurrence of an event. The currently executing event of event system $\symbcore$ in the event context is updated. The $\textsc{EvtSet}$, $\textsc{EvtSeq1}$, and $\textsc{EvtSeq2}$ rules means that when an event occurs in an event set, the event executes until it finishes in the event system. The $\textsc{Par}$ rule shows that execution of a parallel event system is modeled by a nondeterministic interleaving of the atomic execution of event systems. $\symbpes(\symbcore \mapsto \symbevtsys')$ is a function derived from $\symbpes$ by mapping $\symbcore$ to $\symbevtsys'$.

\begin{figure*}
\centering
\footnotesize
\begin{tabular}{ccc}
\infer[\textsc{Basic}]{(\cmdbasic{f}, \symbstate, \symbevtctx) \tran{c} (\cmdnone, f \ \symbstate, \symbevtctx)}{-} \hspace{0.3cm}
&
\infer[\textsc{Seq1}]{(\cmdseq{\symbprog_1}{\symbprog_2}, \symbstate, \symbevtctx) \tran{\symbpcomp} (\symbprog_2, \symbstate', \symbevtctx)}{(\symbprog_1,\symbstate, \symbevtctx) \tran{\symbpcomp} (\cmdnone,\symbstate', \symbevtctx)} \hspace{0.3cm}
&
\infer[\textsc{Seq2}]{(\cmdseq{\symbprog_1}{\symbprog_2}, \symbstate, \symbevtctx) \tran{\symbpcomp} (\cmdseq{\symbprog_1'}{\symbprog_2}, \symbstate', \symbevtctx)}{(\symbprog_1,\symbstate, \symbevtctx) \tran{\symbpcomp} (\symbprog_1',\symbstate', \symbevtctx)} 
\end{tabular}
\vspace{0.3cm}

\begin{tabular}{ccc}
\infer[\textsc{CondT}]{(\cmdcond{\symbbexp}{\symbprog_1}{\symbprog_2}, \symbstate, \symbevtctx) \tran{\symbpcomp} (\symbprog_1, \symbstate, \symbevtctx)}{\symbstate \in \symbbexp} \hspace{0.2cm}
&
\infer[\textsc{CondF}]{(\cmdcond{\symbbexp}{\symbprog_1}{\symbprog_2}, \symbstate, \symbevtctx) \tran{\symbpcomp} (\symbprog_2, \symbstate, \symbevtctx)}{\symbstate \notin \symbbexp} \hspace{0.2cm}
&
\infer[\textsc{WhileF}]{(\cmdwhile{\symbbexp}{\symbprog}), \symbstate, \symbevtctx) \tran{\symbpcomp} (\cmdnone, \symbstate, \symbevtctx)}{\symbstate \notin \symbbexp}
\end{tabular}
\vspace{0.3cm}

\begin{tabular}{ccc}
\infer[\textsc{WhileT}]{(\cmdwhile{\symbbexp}{\symbprog}, \symbstate, \symbevtctx) \tran{\symbpcomp} (\cmdseq{\symbprog}{(\cmdwhile{\symbbexp}{\symbprog})}, \symbstate, \symbevtctx)}{\symbstate \in \symbbexp} \hspace{0.0cm}
&
\infer[\textsc{Await}]{(\cmdawait{\symbbexp}{\symbprog}, \symbstate, \symbevtctx) \tran{\symbpcomp} (\cmdnone, \symbstate', \symbevtctx)}{\symbstate \in \symbbexp & (\symbprog, \symbstate, \symbevtctx) \tran{\symbpcomp^*} (\cmdnone, \symbstate', \symbevtctx)} \hspace{0.0cm}
&
\infer[\textsc{Nondt}]{(\cmdnondt{r}, \symbstate, \symbevtctx) \tran{\symbpcomp} (\cmdnone, \symbstate', \symbevtctx)}{(\symbstate,\symbstate') \in r}
\end{tabular}
\vspace{0.3cm}

\begin{tabular}{cc}
\infer[\textsc{AnonyEvt}]{(\anonevt{\symbprog},\symbstate,\symbevtctx) \trank{\symbpcomp}{k} (\anonevt{\symbprog'},\symbstate',\symbevtctx)}{(\symbprog,\symbstate, \symbevtctx) \tran{\symbpcomp} (\symbprog',\symbstate', \symbevtctx)} \hspace{0.5cm}
&
\infer[\textsc{BasicEvt}]{(\event{\symbevtbd}, \symbstate, \symbevtctx) \trank{\event{\symbevtbd}}{\symbcore} (\anonevt{\symbprog}, \symbstate, \symbevtctx')}{\symbprog = snd(\symbevtbd(p,\symbcore)) & \symbstate \in fst(\alpha(p,\symbcore)) & \symbevtctx' = \symbevtctx(k \mapsto \event{\symbevtbd})}
\end{tabular}
\vspace{0.3cm}

\begin{tabular}{cc}
\infer[\textsc{EvtSet}]{(\evtsys{\symbEvt_0}{\symbEvt_1}{\symbEvt_n}), \symbstate, \symbevtctx) \trank{\symbEvt_i}{\symbcore} (\evtseq{\symbEvt_i'}{(\evtsys{\symbEvt_0}{\symbEvt_1}{\symbEvt_n)},\symbstate, \symbevtctx')}}{i \leq n & (\symbEvt_i, \symbstate, \symbevtctx) \trank{\symbEvt_i}{\symbcore} (\symbEvt_i',\symbstate, \symbevtctx')} \hspace{1cm}
&
\infer[\textsc{EvtSeq1}]
{(\symbEvt \ ; \symbevtsys, \symbstate, \symbevtctx) \trank{\symbact}{\symbcore} (\evtseq{\symbEvt'}
{\symbevtsys}, \symbstate', \symbevtctx)}{(\symbEvt, \symbstate, \symbevtctx) \trank{\symbact}{\symbcore} (\symbEvt', \symbstate', \symbevtctx)}
\end{tabular}
\vspace{0.3cm}

\begin{tabular}{cc}
\infer[\textsc{EvtSeq2}]
{(\symbEvt \ ; \symbevtsys, \symbstate, \symbevtctx) \trank{\symbact}{\symbcore} (\symbevtsys, \symbstate', \symbevtctx)}
{(\symbEvt, \symbstate, \symbevtctx) \trank{\symbact}{\symbcore} (\anonevt{\cmdnone}, \symbstate', \symbevtctx)} \hspace{1cm}
&
\infer[\textsc{Par}]{(\symbpes, \symbstate, \symbevtctx) \trank{\symbact}{\symbcore} (\symbpes', \symbstate', \symbevtctx')}{(\symbpes(\symbcore), \symbstate, \symbevtctx) \trank{\symbact}{\symbcore} (\symbevtsys', \symbstate', \symbevtctx') & \symbpes' = \symbpes(\symbcore \mapsto \symbevtsys')}
\end{tabular}
\caption{Operational Semantics of {\slang} Language}
\label{fig:semantics}
\end{figure*}

\subsection{Computation}

A \emph{computation} of {\slang} is a sequence of transitions, which is defined as the form
\[
\symbconf_0 \tran{\symbtran_0} \symbconf_1 \tran{\symbtran_1}  ... \tran{\symbtran_n} \symbconf_n \tran{\symbtran_{n+1}} ... , (where \ \symbtran ::= \symbactk \mid e)
\]

We define the set of computations of parallel event systems $\compstps$, as the set of lists of configurations inductively defined as follows, where $\#$ is the connection operator of two lists. The one-element list of configurations is always a computation. Two consecutive configurations are part of a computation if they are the initial and final configurations of an environment or action transition. 

\begin{equation*}
\left\{
\begin{aligned}
& [(\symbpes, \symbstate, \symbevtctx)] \in  \compstps \\
& (\symbpes, \symbstate_1, \symbevtctx_1)\#cs \in  \compstps \\
& \quad \quad \quad \Longrightarrow (\symbpes, \symbstate_2, \symbevtctx_2)\#(\symbpes, \symbstate_1, \symbevtctx_1)\#cs \in  \compstps \\
& (\symbpes_2, \symbstate_2, \symbevtctx_2) \tran{\symbactk} (\symbpes_1, \symbstate_1, \symbevtctx_1) \wedge (\symbpes_1, \symbstate_1, \symbevtctx_1)\#cs \in  \compstps \\
& \quad \quad \quad \Longrightarrow (\symbpes_2, \symbstate_2, \symbevtctx_2)\#(\symbpes_1, \symbstate_1, \symbevtctx_1)\#cs \in  \compstps
\end{aligned}
\right.
\end{equation*}

The computations of programs, events, and event systems are defined in a similar way. We use $\compfun(\symbpes)$ to denote the set of computations of a parallel event system $\symbpes$. The function $\compfun(\symbpes, \symbstate, \symbevtctx)$ denotes the computations of $\symbpes$ executing from an initial state $\symbstate$ and event context $\symbevtctx$. The computations of programs, events, and event systems are also denoted as the $\compfun$ function. For each computation $\symbcomp \in \compfun(\symbpes)$, we use $\symbcomp_i$ to denote the configuration at index $i$. For convenience, we use $\symbcomp$ to denote computations of programs, events, and event systems too.
We say that a parallel event system $\symbpes$ is a \emph{closed system} when there is no environment transition in computations of $\symbpes$. 

We define an equivalent relation on computations as follows. Here, we concern the state, event context, and transitions, but not the specification of a configuration. 
\begin{definition}[Simulation of Computations]
A computation $\symbcomp_1$ is a simulation of $\symbcomp_2$, denoted as $\compsim{\symbcomp_1}{\symbcomp_2}$, if 
\begin{itemize}
\item $len(\symbcomp_1) = len(\symbcomp_2)$
\item $\forall i < len(\symbcomp_1) - 1. \ \symbstate_{\symbcomp_{1_i}} = \symbstate_{\symbcomp_{2_i}} \wedge \symbevtctx_{\symbcomp_{1_i}} = \symbevtctx_{\symbcomp_{2_i}} \wedge (\symbcomp_{1_i} \tran{\symbactk} \symbcomp_{1_{i+1}}) = (\symbcomp_{2_i} \tran{\symbactk} \symbcomp_{2_{i+1}})$
\end{itemize}
\end{definition}

\section{Information-flow Security of {\slang}}
\label{sect:ifs}
This section discusses state-action based IFS of the {\slang} language. We consider the security of parallel event systems that are closed. We first introduce the security policies. Then, we construct a state machine from {\slang}. Based on the state machine, we present the security properties and the unwinding theorem. 

\subsection{IFS Configuration}
In order to discuss the security of a parallel event system $\symbpes$, we assume a set of security domains $\symbDomain$ and a security policy $\interf$ that restricts the allowable flow of information among those domains. The security policy $\interf$ is a reflexive relation on $\symbDomain$. 
$\symbdomain_1 \interf \symbdomain_2$ means that actions performed by $\symbdomain_1$ can influence subsequent outputs seen by $\symbdomain_2$. $\ninterf$ is the complement relation of $\interf$. 
We call $\interf$ and $\ninterf$ the \emph{interference} and \emph{noninterference} relations respectively.

Each event has an execution domain. Traditional formulations in the state-action based IFS assume a static mapping from events to domains, such that the domain of an event can be determined solely from the event itself \cite{rushby92,Oheimb04}. For flexibility, we use a dynamic mapping, which is represented by a function $dom\_e: S \times \symbCore \times \symbEvt \rightarrow \symbDomain$, where $S$ is the system state.
The $\symbpes$ is \emph{view-partitioned} if, for each domain $\symbdomain \in \symbDomain$, there is an equivalence relation $\dsim{\symbdomain}$ on $\symbState$. For convenience, we define $\symbconf_1 \dsim{\symbdomain} \symbconf_2 \defi \symbstate_{\symbconf_1} \dsim{\symbdomain} \symbstate_{\symbconf_2}$. 
An observation function of a domain $\symbdomain$ to a state $\symbstate$ is defined as $ob(\symbstate,\symbdomain)$. For convenience, we define $ob(\symbconf,\symbdomain) \defi ob(\symbstate_\symbconf,\symbdomain)$.

\subsection{State Machine Representation of {\slang}}
IFS semantics of {\slang} consider small-step actions of systems. A small-step action in the machine is identified by the label of a transition, the event that the action belongs to, and the domain that triggers the event. 
We construct a nondeterministic state machine for a parallel event system as follows. 

\begin{definition}
\label{def:statemachine}
A state machine of a closed $\symbpes$ executing from an initial state $\symbstate_0$ and initial event context $\symbevtctx_0$ is a quadruple $\symbSM=\langle \symbConf, \symbAction, step, \symbconf_0 \rangle$, where 
\begin{itemize}
\item $\symbConf$ is the set of configurations.
\item $\symbAction$ is the set of actions. An action is a triple $\symbaction = \langle \symbactk, \symbevt, \symbdomain \rangle$, where $\symbactk$ is a transition label, $\symbevt$ is an event, and $\symbdomain$ is a domain. 
\item $step: \symbAction \rightarrow \mathbb{P}(\symbConf \times \symbConf)$ is the transition function, where $step(\symbaction) = \{(\symbconf,\symbconf') \mid \symbconf \tran{\symbactk_\symbaction} \symbconf' \wedge ((\symbactk_\symbaction = \actk{\symbevt_\symbaction}{\symbcore} \wedge dom\_e(\symbstate_\symbconf, \symbcore, \symbevt_\symbaction) = \symbdomain_\symbaction) \vee (\symbactk_\symbaction = \actk{\symbpcomp}{\symbcore} \wedge \symbevt_\symbaction = \symbevtctx_\symbconf(\symbcore) \wedge dom\_e(\symbstate_\symbconf, \symbcore, \symbevt_\symbaction) = \symbdomain_\symbaction))\}$. 
\item $\symbconf_0 = \langle \symbpes, \symbstate_0, \symbevtctx_0 \rangle$ is the initial configuration.
\end{itemize}
\end{definition}

Based on the function $step$, we define the function $run$ as shown in {\figprefix} \ref{fig:aux_funs} to represent the execution of a sequence of actions. We prove the following lemma to ensure that the state machine is an equivalent representation of the {\slang} language. 

\begin{lemma}
\label{lm:sm_equiv}
The state machine defined in {\defprefix} \ref{def:statemachine} is an equivalent representation of {\slang}, i.e., 
\begin{itemize}
\item If $(\symbconf_1,\symbconf_2) \in run(\symbactions)$, then $\exists \symbcomp. \ \symbcomp \in \compstps \wedge \symbcomp_0 = \symbconf_1 \wedge last(\symbcomp) = \symbconf_2 \wedge (\forall j < len(\symbcomp) - 1. \ \symbcomp_j \tran{\symbactk_{\symbactions_j}} \symbcomp_{(j+1)})$, and 
\item If $\symbcomp \in \compstps \wedge \symbcomp_0 = \symbconf_1 \wedge last(\symbcomp) = \symbconf_2 \wedge (\forall j < len(\symbcomp) - 1. \ \neg (\symbcomp_j \evtran \symbcomp_{(j+1)}))$, then $\exists \symbactions. \ (\symbconf_1,\symbconf_2) \in run(\symbactions) \wedge (\forall j < len(\symbcomp) - 1. \ \symbcomp_j \tran{\symbactk_{\symbactions_j}} \symbcomp_{(j+1)})$
\end{itemize}
\end{lemma}

Since we consider closed parallel event systems, there is no environment transition in the computations of $\symbpes$, i.e., $\forall j < len(\symbcomp) - 1. \ \neg (\symbcomp_j \evtran \symbcomp_{(j+1)})$.

\subsection{Information-flow Security Properties}
We now discuss the IFS properties based on the state machine constructed above. 
By following the security properties in \cite{Oheimb04}, we define \emph{noninterference}, \emph{nonleakage}, and \emph{noninfluence} properties in this work. 

The auxiliary functions used by IFS are defined in detail in {\figprefix} \ref{fig:aux_funs}. 
The function $execution(\symbconf,as)$ (denoted as $\execution{\symbconf}{as}$) returns the set of final configurations by executing a sequence of actions $as$ from a configuration $\symbconf$, where $\lhd$ is the domain restriction of a relation. By the function $execution$, the reachability of a configuration $\symbconf$ from the initial configuration $\symbconf_0$ is defined as $reachable(\symbconf)$ (denoted as $\reachablef(\symbconf)$). 

\begin{figure}[t]
{
\footnotesize
\begin{tabular}{ll} 
$ 
\left\{
\begin{aligned}
& run([\ ]) = Id \\
& run(\symbaction \# \symbactions) = step(\symbaction) \circ run(\symbactions)
\end{aligned}
\right.
$ 
&
$ 
\begin{aligned}
& \execution{\symbconf}{\symbactions} \defi \{\symbconf\} \lhd run(\symbactions) \\
\end{aligned}
$ 
\vspace{5pt}
\\
\multicolumn{2}{l}{
$\equidoms{\symbactions_1}{\symbdoms}{\symbactions_2} \defi \forall \symbdomain \in \symbdoms.\ \equidom{\symbconf_1}{d}{\symbconf_2}$
\quad \quad
$\reachablef(\symbconf) \defi  \exists \symbactions. \ (\symbconf_0,\symbconf)\in run(\symbactions)$
}
\vspace{5pt}
\\ 
\multicolumn{2}{l}{
$\ssequidom{\symbconf s_1}{\symbdomain}{\symbconf s_2} \defi \forall \symbconf_1 \ \symbconf_2. \ \symbconf_1 \in \symbconf s_1 \wedge \symbconf_2 \in \symbconf s_2 \longrightarrow ob(\symbconf_1,d) = ob(\symbconf_2,d)$
}
\vspace{5pt}
\\ 
\multicolumn{2}{l}{
$ 
\left\{
\begin{aligned}
& sources ([ \ ], \symbdomain) = \{\symbdomain\} \\
& sources (\symbaction \# \symbactions, \symbdomain) = 
\begin{aligned}
& sources (\symbactions, \symbdomain) \cup \{w.\ w = \symbdomain_\symbaction \wedge \\
& (\exists v. \ (w \interf v) \wedge v \in sources(\symbactions, \symbdomain))\}
\end{aligned}
\end{aligned}
\right.
$ 
}
\vspace{5pt}
\\
\multicolumn{2}{l}{
$ 
\left\{
\begin{aligned}
& ipurge ([ \ ], \symbdomain) = [ \ ] \\
& ipurge (\symbaction \# \symbactions, \symbdomain) = 
\begin{aligned}
& \mathbf{if} \ \symbdomain_\symbaction \in sources(\symbaction \# \symbactions, \symbdomain) \ \mathbf{then} \\
& \quad \symbaction \# ipurge (as, \symbdomain) \\
& \mathbf{else} \ \ ipurge (\symbactions, \symbdomain) 
\end{aligned}
\end{aligned}
\right.
$ 
}
\end{tabular}
}
\caption{Auxiliary Functions of Information-flow Security}
\label{fig:aux_funs}
\end{figure}

The essence of intransitive noninterference is that a domain $d$ cannot distinguish the final states between executing a sequence of actions $as$ and executing its purged sequence. In the intransitive purged sequence ($ipurge(\symbactions,\symbdomain)$ in {\figprefix} \ref{fig:aux_funs}), the actions of domains that are not allowed to pass information to $d$ directly or indirectly are removed.
In order to express the allowed information flows for the intransitive policies, we use a function $sources(\symbactions,\symbdomain)$ as shown in {\figprefix} \ref{fig:aux_funs}, which yields the set of domains that are allowed to pass information to a domain $\symbdomain$ when an action sequence $\symbactions$ executes.

The observational equivalence of an execution is thus denoted as  $\equivexec{\symbconf_1}{\symbactions_1}{\symbdomain}{\symbconf_2}{\symbactions_2}$, which means that a domain $\symbdomain$ is identical to any two final states after executing $\symbactions_1$ from $\symbconf_1$ ($\execution{\symbconf_1}{\symbactions_1}$) and executing $\symbactions_2$ from $\symbconf_2$. The classical nontransitive noninterference \cite{rushby92} is defined as the \emph{noninterference} property as follows. 

\begin{equation*}
\begin{aligned}
noninterference \defi \forall \ \symbactions, \symbdomain. \ \equivexec{\symbconf_0}{\symbactions}{\symbdomain}{\symbconf_0}{ipurge(\symbactions,\symbdomain)}
\end{aligned}
\end{equation*}

The above definition of noninterference is based on the initial configuration $\symbconf_0$, but concurrent systems usually support \emph{warm} or \emph{cold start} and they may start to execute from a non-initial configuration. Therefore, we define a more general version \emph{noninterferece\_r} as follows based on the function $reachable$. This general noninterference requires that the system starting from any reachable configuration is secure. It is obvious that this noninterference implies the classical noninterference due to $\reachablef(\symbconf_0) = True$.

\begin{equation*}
\begin{aligned}
nonint&erference\_r \defi \\
& \forall \ \symbactions, \symbdomain, \symbconf. \ \reachablef(\symbconf)
\longrightarrow \equivexec{\symbconf}{\symbactions}{\symbdomain}{\symbconf}{ipurge(\symbactions,\symbdomain)}
\end{aligned}
\end{equation*}

The intuitive meaning of \emph{nonleakage} is that if data are not leaked initially, data should not be leaked during executing a sequence of actions. Concurrent systems are said to preserve nonleakage when for any pair of reachable configuration $\symbconf_1$ and $\symbconf_2$ and an observing domain $d$, if (1) $\symbconf_1$ and $\symbconf_2$ are equivalent for all domains that may (directly or indirectly) interfere with $d$ during the execution of $\symbactions$, i.e. $\equidoms{\symbconf_1}{sources(\symbactions, \symbdomain)}{\symbconf_2}$, then $\symbconf_1$ and $\symbconf_2$ are observationally equivalent for $\symbdomain$ and $\symbactions$. Noninfluence is the combination of nonleakage and classical noninterference. Noninfluence ensures that there is no secrete data leakage and secrete actions are not visible according to the information-flow security policies. The two security properties are defined as follows. We have that \emph{noninfluence} implies \emph{noninterference\_r}.

\begin{equation*}
\begin{aligned}
nonleakage \defi & \forall \ \symbconf_1, \symbconf_2, \symbdomain, \symbactions. \ \reachablef(\symbconf_1) \wedge \reachablef(\symbconf_2) \\
& \longrightarrow \equidoms{\symbconf_1}{sources(\symbactions, \symbdomain)}{\symbconf_2} \longrightarrow \equivexec{\symbconf_1}{\symbactions}{\symbdomain}{\symbconf_2}{\symbactions}
\end{aligned}
\end{equation*}

\begin{equation*}
\begin{aligned}
non&influence \defi \forall \ \symbconf_1, \symbconf_2, \symbdomain, \symbactions. \ \reachablef(\symbconf_1) \wedge \reachablef(\symbconf_2) \\
& \longrightarrow \equidoms{\symbconf_1}{sources(\symbactions, \symbdomain)}{\symbconf_2} \longrightarrow \equivexec{\symbconf_1}{\symbactions}{\symbdomain}{\symbconf_2}{ipurge(\symbactions,\symbdomain)}
\end{aligned}
\end{equation*}


%
%
%
%
%

\subsection{Small-step Unwinding Conditions and Theorem}
\label{subsec:small_ucs}


The standard proof of IFS is discharged by proving a set of unwinding conditions \cite{rushby92} that examine individual execution steps of the system. This paper also follows this approach. We first define the small-step unwinding conditions as follows.

\begin{definition}[Observation Consistent - OC]
For a parallel event system $\symbpes$, the equivalence relation $\dsim{}$ are said to be \emph{observation consistent} if
\begin{equation*}
\forall \symbconf_1, \symbconf_2, \symbdomain. \ \symbconf_1 \dsim{\symbdomain} \symbconf_2 \longrightarrow ob(\symbconf_1, \symbdomain) = ob(\symbconf_2, \symbdomain)
\end{equation*}
\end{definition}

\begin{definition}[Locally Respects - LR]
A parallel event system $\symbpes$ locally respects $\interf$ if 
\begin{equation*}
\begin{aligned}
\forall \symbaction, \symbdomain, \symbconf. \ & \reachablef(\symbconf) \longrightarrow  
\symbdomain_\symbaction \ninterf \symbdomain \longrightarrow \\
& (\forall \symbconf'. \ (\symbconf, \symbconf') \in step(\symbaction) \longrightarrow \symbconf \dsim{\symbdomain} \symbconf')
\end{aligned}
\end{equation*}
\end{definition}

\begin{definition}[Step Consistent - SC]
A parallel event system $\mathcal{PS}$ is step consistent if 
\begin{equation*}
\begin{aligned}
\forall \symbaction, \symbdomain, & \symbconf_1, \symbconf_2. \ \reachablef(\symbconf_1) \wedge \reachablef(\symbconf_2) \longrightarrow \\
&\symbconf_1 \dsim{\symbdomain} \symbconf_2 
\wedge ((\symbdomain_\symbaction \interf \symbdomain) \longrightarrow (\symbconf_1 \dsim{\symbdomain_\symbaction} \symbconf_2)) \longrightarrow \\
&
\begin{aligned}
(\forall \symbconf_1', \symbconf_2'. \ (\symbconf_1, \symbconf_1') \in step(\symbaction) \wedge (\symbconf_2, \symbconf_2') \in step(\symbaction) \\
\longrightarrow \symbconf_1' \dsim{\symbdomain} \symbconf_2')
\end{aligned}
\end{aligned}
\end{equation*}
\end{definition}

The locally respects condition means that an action $\symbaction$ that executes in a configuration $\symbconf$ can affect only those
domains to which the domain executing $\symbaction$ is allowed to send information. The step consistent condition says that the observation by a domain $\symbdomain$ after an action $\symbaction$ occurs can depend only on $\symbdomain$'s observation before $\symbaction$ occurs, as well as the observation by the domain executing $\symbaction$ before $\symbaction$ occurs if that domain is allowed to send information to $\symbdomain$. 

We prove the small-step unwinding theorem for \emph{noninfluence} and \emph{nonleakage} as follows. 
\begin{theorem}[Unwinding Theorem of Noninfluence]
\label{thm:ut_noninfl}
\[ OC \wedge LR \wedge SC \Longrightarrow noninfluence \]
\end{theorem}

\begin{theorem}[Unwinding Theorem of Nonleakage]
\label{thm:ut_nonlk}
\[ OC \wedge LR \wedge SC \Longrightarrow nonleakage \]
\end{theorem}

%
%
%
%
%

\section{Rely-Guarantee Proof System for {\slang}}
\label{sect:rgproof}
For the purpose of compositional reasoning of IFS, we propose a rely-guarantee proof system for {\slang} in this section. 
We first introduce the rely-guarantee specification and its validity. Then, a set of proof rules and their soundness for the compositionality are discussed.

\subsection{Rely-Guarantee Specification}

A rely-guarantee specification for a system is a quadruple $RGCond = \rgconddefault$, where $pre$ is the pre-condition, $R$ is the rely condition, $G$ is the guarantee condition, and $pst$ is the post condition. The assumption and commitment functions following a standard way are defined as follows. 

\begin{equation*}
\begin{aligned}
\assumefun(pre, R) \defi \{\symbcomp \mid \ & \symbstate_{\symbcomp_0} \in pre \wedge (\forall i < len(\symbcomp) - 1. \\
& (\symbcomp_i \evtran \symbcomp_{i+1}) \longrightarrow (\symbstate_{\symbcomp_i},\symbstate_{\symbcomp_{i+1}}) \in R)\}
\end{aligned}
\end{equation*}

\begin{equation*}
\begin{aligned}
\commitfun(G, pst) \defi \{\symbcomp \mid \ & (\forall i < len(\symbcomp) - 1. \\
& (\symbcomp_i \tran{\symbactk} \symbcomp_{i+1}) \longrightarrow (\symbstate_{\symbcomp_i},\symbstate_{\symbcomp_{i+1}}) \in G) \\
& \wedge (\symbspec_{last(\symbcomp)} = \cmdnone \longrightarrow \symbstate_{\symbcomp_n} \in pst)\}
\end{aligned}
\end{equation*}

For an event, the commitment function is similar, but the condition $\symbspec_{last(\symbcomp)} = \anonevt{\cmdnone}$. Since event systems and parallel event systems execute forever, the commitment function of them is defined as follows. We release the condition on the final state. 

\begin{equation*}
\begin{aligned}
\commitfun(G, pst) \defi \{\symbcomp \mid \ & (\forall i < len(\symbcomp) - 1. \\
& (\symbcomp_i \tran{\symbactk} \symbcomp_{i+1}) \longrightarrow (\symbstate_{\symbcomp_i},\symbstate_{\symbcomp_{i+1}}) \in G)\}
\end{aligned}
\end{equation*}

Validity of rely-guarantee specification in a parallel event system means that the system satisfies the specification, which is precisely defined as follows. Validity for programs, events, and event systems are defined in a similar way.

\begin{definition}[Validity of Rely-Guarantee Specification]
A parallel event system $\symbpes$ satisfies its specification $\rgconddefault$, denoted as $\RGSAT{\symbpes}{\rgcond{pre}{R}{G}{pst}}$, iff $\forall \symbstate, \symbevtctx. \ \compfun(\symbpes, \symbstate, \symbevtctx) \cap \assumefun(pre, R) \subseteq \commitfun(G, pst)$.
\end{definition}

\subsection{Proof Rules}
We present the proof rules in {\figprefix} \ref{fig:proofrule}, which gives us a relational proof method for concurrent systems. $UNIV$ is the universal set.

\begin{figure*}
\centering
\footnotesize
\begin{tabular}{cc}
\infer[\textsc{Basic}]{\rgsat{(\cmdbasic{f})}{\rgconddefault}}
{
\begin{tabular}{l}
$pre \subseteq \{\symbstate \mid f(\symbstate) \in pst\} \quad stable(pre, R) \quad stable(pst, R)$ \\ 
$\{(\symbstate, \symbstate') \mid \symbstate \in pre \wedge (\symbstate' = f(\symbstate) \vee \symbstate' = \symbstate)\} \in G$
\end{tabular}
}\hspace{1cm}
&
\infer[\textsc{Seq}]{\rgsat{(\cmdseq{P}{Q})}{\rgconddefault}}
{
\rgsat{P}{\rgcond{pre}{R}{G}{m}} & \rgsat{Q}{\rgcond{m}{R}{G}{pst}} 
}
\end{tabular}
\vspace{0.3cm}

\begin{tabular}{cc}
\infer[\textsc{Cond}]{\rgsat{(\cmdcond{\symbbexp}{P_1}{P_2})}{\rgconddefault}}
{
\begin{tabular}{l}
$stable(pre, R) \quad \rgsat{P_1}{\rgcond{pre \cap \symbbexp}{R}{G}{pst}}$ \\ 
$\rgsat{P_2}{\rgcond{pre \cap - \symbbexp}{R}{G}{pst}} \quad \forall \symbstate. \ (\symbstate, \symbstate) \in G$
\end{tabular}
}\hspace{1cm}
&
\infer[\textsc{While}]{\rgsat{(\cmdwhile{\symbbexp}{P})}{\rgconddefault}}
{
\begin{tabular}{l}
$stable(pre, R) \quad pre \cap - \symbbexp \subseteq pst \quad stable(pst, R)$ \\ 
$\rgsat{P}{\rgcond{pre \cap \symbbexp}{R}{G}{pst}} \quad \forall \symbstate. \ (\symbstate, \symbstate) \in G$
\end{tabular}
}
\end{tabular}
\vspace{0.3cm}

\begin{tabular}{c}
\infer[\textsc{Await}]{\rgsat{(\cmdawait{\symbbexp}{P})}{\rgconddefault}}
{
\begin{tabular}{l}
$\forall V.\ \rgsat{P}{\rgcond{pre \cap \symbbexp \cap {V}}{\{(\symbstate, \symbstate') \mid \symbstate = \symbstate'\}}{UNIV}{\{\symbstate \mid (V, \symbstate) \in G\} \cap pst}} \quad stable(pre, R) \quad stable(pst, R)$
\end{tabular}
}
\end{tabular}
\vspace{0.3cm}

\begin{tabular}{cc}
\infer[\textsc{Nondt}]{\rgsat{(\cmdnondt{r})}{\rgconddefault}}
{
\begin{tabular}{l}
$pre \subseteq \{s. \ (\forall \symbstate'. \ (\symbstate, \symbstate') \in r \longrightarrow \symbstate' \in pst) \wedge (\exists \symbstate'. \ (\symbstate, \symbstate') \in r)\} $
\\
$\{(\symbstate, \symbstate') \mid \symbstate \in pre \wedge (\symbstate, \symbstate') \in r\} \subseteq G \quad stable(pre, R) \quad stable(pst, R)$
\end{tabular}
}\hspace{0.5cm}
&
\infer[\textsc{AnonyEvt}]{\rgsat{(\anonevt{P})}{\rgconddefault}}
{\rgsat{P}{\rgconddefault}}
\end{tabular}
\vspace{0.3cm}

\begin{tabular}{cc}
\infer[\textsc{BasicEvt}]{\rgsat{\event{\symbevtbd}}{\rgconddefault}}
{
\begin{tabular}{l}
$\forall p, \symbcore. \ \rgsat{snd(\symbevtbd(p, \symbcore))}{\rgcond{pre \cap fst(\symbevtbd(p, \symbcore))}{R}{G}{pst}}$\\
$stable(pre, R) \quad \forall \symbstate. \ (\symbstate, \symbstate) \in G$
\end{tabular}
}\hspace{0.5cm}
&
\infer[\textsc{EvtSeq}]{\rgsat{(\evtseq{\symbEvt}{\symbevtsys})}{\rgconddefault}}
{
\begin{tabular}{l}
$\rgsat{\symbEvt}{\rgcond{pre}{R}{G}{m}} \quad \rgsat{\symbevtsys}{\rgcond{m}{R}{G}{pst}}$ 
\end{tabular}
}
\end{tabular}
\vspace{0.3cm}

\begin{tabular}{cc}
\infer[\textsc{EvtSet}]{\rgsat{(\evtsys{\symbEvt_0}{\symbEvt_1}{\symbEvt_n})}{\rgconddefault}}
{
\begin{tabular}{l}
$\forall i \leq n. \ \rgsat{\symbEvt_i}{\rgcond{pres_i}{Rs_i}{Gs_i}{psts_i}} \quad \forall i \leq n. \ pre \subseteq pres_i$ \\
$\forall i \leq n. \ R \subseteq Rs_i \quad \forall i \leq n. \ Gs_i \subseteq G \quad \forall i \leq n. \ psts_i \subseteq pst$ \\ 
$\forall i \leq n, j \leq n. \ psts_i \subseteq pres_j \quad stable(pre, R) \quad \forall \symbstate. \ (\symbstate, \symbstate) \in G$
\end{tabular}
}\hspace{0.5cm}
&
\infer[\textsc{Conseq}]{\rgsat{\symbspec}{\rgconddefault}}
{
\begin{tabular}{l}
$pre \subseteq pre' \quad R \subseteq R' \quad G' \subseteq G \quad pst' \subseteq pst$
\\
$\quad \rgsat{\symbspec}{\rgcond{pre'}{R'}{G'}{pst'}}$
\end{tabular}
}
\end{tabular}
\vspace{0.3cm}

\begin{tabular}{c}
\infer[\textsc{Par}]{\rgsat{\symbpes}{\rgconddefault}}
{
\begin{tabular}{l}
$\symbpes = (\parsysc) \quad \forall \symbcore. \ \rgsat{\symbpes(\symbcore)}{\rgcond{pres_\symbcore}{Rs_\symbcore}{Gs_\symbcore}{psts_\symbcore}} \quad \forall \symbcore. \ pre \subseteq pres_\symbcore$ \\
$\forall \symbcore. \ R \subseteq Rs_\symbcore \quad \forall \symbcore. \ Gs_\symbcore \subseteq G \quad \forall \symbcore. \ psts_\symbcore \subseteq pst \quad \forall \symbcore, \symbcore'. \ \symbcore \neq \symbcore' \longrightarrow Gs_\symbcore \subseteq Rs_{\symbcore'}$
\end{tabular}
}
\end{tabular}

\caption{Rely-guarantee Proof Rules for {\slang}}
\label{fig:proofrule}
\end{figure*}

The proof rules for programs are mostly standard \cite{Xu97, Nieto03}. For $\cmdnondt{r}$, any state change in $r$ requires that $pst$ holds immediately after the action transition and the transition should be in $G$ relation. Before and after this action transition there may be a number of environment transitions, $stable(pre, R)$ and $stable(pst,R)$ ensure that $pre$ and $pst$ hold during any number of environment transitions in $R$ before and after the action transition, respectively. 

An anonymous event is just a wrapper of a program, and they have the same state and event context in their computations according to the $\textsc{AnonyEvt}$ transition rule in {\figprefix} \ref{fig:semantics}. Therefore, $\anonevt{P}$ satisfies the rely-guarantee specification iff the program $P$ satisfies the specification. A basic event is actually a parametrized program with a list of parameters $p$ and a execution context $\symbcore$. A basic event satisfies its rely-guarantee specification, if for any program mapping from $p$ and $\symbcore$ satisfies the rely-guarantee condition with augmented pre-condition by the guard condition of the event. Since the occurrence of an event does not change the state ($\textsc{BasicEvt}$ rule in {\figprefix} \ref{fig:semantics}), we require that $\forall \symbstate. \ (\symbstate, \symbstate) \in G$. Moreover, there may be a number of environment transitions before the event occurs. $stable(pre, R)$ ensures that $pre$ holds during the environment transitions. 

We now introduce the proof rule for event systems. The $\textsc{EvtSeq}$ rule is similar to $\textsc{Seq}$ and is intuitive. 
Recall that when an event occurs in an event set, the event executes until it finishes in the event system. Then, the event system behaves as the event set. Thus, events in an event system do not execute in interleaving manner. To prove that an event set holds its rely-guarantee specification $\rgconddefault$, we have to prove eight premises ($\textsc{EvtSet}$ rule in {\figprefix} \ref{fig:proofrule}). The first one requires that each event together with its specification be derivable in the system. The second one requires that the pre-condition for the event set implies all the event's preconditions. The third one is a constraint on the rely condition of event $i$. An environment transition for $i$ corresponds to a transition from the environment of the event set. The fourth one imposes a relation among the guarantee conditions of events and that of the event set. Since an action transition of the event set is performed by one of its events, the guarantee condition $Gs_i$ of each event must be in the guarantee condition of the event set. The fifth one requires that the post-condition of each event must be in the overall post-condition. Since the event set behaves as itself after an event finishes, the sixth premise says that the post-condition of each event should imply the pre-condition of each event. The meaning of the last two premises are the same as we mentioned before. 

The $\textsc{Conseq}$ rule allows us to strengthen the assumptions and weaken the commitments. The meaning of the $\textsc{Par}$ rule is also standard. 

\subsection{Soundness}

The soundness of rules for events is straightforward and is based on the rules for programs, which are proved by the same way in \cite{Xu97}.

To prove soundness of rules for event systems. First, we show how to decompose a computation of event systems into computations of its events. 

\begin{definition}[Serialization of Events]
A computation $\symbcomp$ of event systems is a serialization of a set of events $\{\symbEvt_1, \symbEvt_2, ..., \symbEvt_n\}$, denoted by $\Serialize{\symbcomp}{\{\symbEvt_1, \symbEvt_2, ..., \symbEvt_n\}}$, iff there exist a set of computations $\symbcomp_1, ..., \symbcomp_m$, where for $1 \leq i \leq m$ there exists $1 \leq k \leq n$ that $\symbcomp_i \in \compste(\symbEvt_k)$, such that $\compsim{\symbcomp}{\symbcomp_1 \# \symbcomp_2 \# ... \# \symbcomp_m}$.
\end{definition}

\begin{lemma}
\label{lm:seri}
For any computation $\symbcomp$ of an event system $\symbevtsys$, $\Serialize{\symbcomp}{evts(\symbevtsys)}$. 
\end{lemma}

The soundness of the $\textsc{EvtSeq}$ rule is proved by two cases. For any computation $\symbcomp$ of ``$\evtseq{\symbEvt}{\symbevtsys}$'', the first case is that the execution of event $\symbEvt$ does not finish in $\symbcomp$. In such a case, $\Serialize{\symbcomp}{\{\symbEvt\}}$. By the first premise of this rule, we can prove the soundness; In the second case, the execution of event $\symbEvt$ finishes in $\symbcomp$. In such a case, we have $\symbcomp = \symbcomp_1 \# \symbcomp_2$, where $\Serialize{\symbcomp_1}{\{\symbEvt\}}$ and $\Serialize{\symbcomp_2}{evts(\symbevtsys)}$. By the two premises of this rule, we can prove the soundness. 

The soundness of the $\textsc{EvtSet}$ rule is complicated. 
From {\lemmaprefix} \ref{lm:seri}, we have that for any computation $\symbcomp$ of the event set, $\compsim{\symbcomp}{\symbcomp_1 \# \symbcomp_2 \# ... \# \symbcomp_m}$, for $1 \leq i \leq m$ there exists $1 \leq k \leq n$ that $\symbcomp_i \in \compste(\symbEvt_k)$. 
When $\symbcomp$ is in $\assumefun(pre, R)$, from $\forall i \leq n, j \leq n. \ psts_i \subseteq pres_j$, $\forall i \leq n. \ pre \subseteq pres_i$, and $\forall i \leq n. \ R \subseteq Rs_i$, we have that there is one $k$ for each $\symbcomp_i$ that $\symbcomp_i$ is in $\assumefun(pres_k, Rs_k)$. By the first premise in the $\textsc{EvtSet}$ rule, we have $\symbcomp_i$ is in $\commitfun(Gs_k, psts_k)$. Finally, with $\forall i \leq n. \ Gs_i \subseteq G$ and $\forall i \leq n. \ psts_i \subseteq pst$, we have that $\symbcomp$ is in $\commitfun(G, pst)$. 

Finally, the soundness theorem of the rule for parallel composition is shown as follows. 

\begin{theorem}[\rm Soundness of Parallel Composition Rule]
\label{thm:snd_par}
\[
\rgsat{\symbpes}{\rgconddefault} \Longrightarrow \RGSAT{\symbpes}{\rgconddefault}
\]
\end{theorem}

To prove this theorem, we first use \emph{conjoin} of computations to decompose a computation of parallel event systems into computations of its event systems. 

\begin{definition}
A computation $\symbcomp$ of a parallel event system $\symbpes$ and a set of computations $\symbcompk: \symbCore \rightarrow \compstes$ conjoin, denoted by $\compconjoin{\symbcomp}{\symbcompk}$, iff
\begin{itemize}
\item $\forall \symbcore. \ len(\symbcomp) = len(\symbcompk(\symbcore))$. 
\item $\forall \symbcore, j < len(\symbcomp). \ \symbstate_{\symbcomp_j} = \symbstate_{\symbcompk(\symbcore)_j} \wedge \symbevtctx_{\symbcomp_j} = \symbevtctx_{\symbcompk(\symbcore)_j}$.
\item $\forall \symbcore, j < len(\symbcomp). \ \symbspec_{\symbcomp_j}(\symbcore) = \symbspec_{\symbcompk(\symbcore)_j}$. 
\item for $j < len(\symbcomp) - 1$, one of the following two cases holds:
	\begin{itemize}
	\item $\symbcomp_j \evtran \symbcomp_{j+1}$, and $\forall \symbcore. \ \symbcompk(\symbcore)_j \evtran \symbcompk(\symbcore)_{j+1}$. 
	\item $\symbcomp_j \trank{\symbact}{\symbcore_1} \symbcomp_{j+1}$, $\symbcompk(\symbcore_1)_j \trank{\symbact}{\symbcore_1} \symbcompk(\symbcore_1)_{j+1}$, and $\forall \symbcore \neq \symbcore_1. \ \symbcompk(\symbcore)_j \evtran \symbcompk(\symbcore)_{j+1}$. 
	\end{itemize}
\end{itemize}
\end{definition}

\begin{lemma}
\label{lm:semaitc_compositional}
The semantics of {\slang} is compositional, i.e., $\compfun(\symbpes, \symbstate, \symbevtctx) = \{\symbcomp \mid (\exists \symbcompk \mid (\forall \symbcore. \ \symbcompk(\symbcore) \in \compfun(\symbpes(\symbcore), \symbstate, \symbevtctx)) \wedge \compconjoin{\symbcomp}{\symbcompk})\}$. 
\end{lemma}

The soundness of the $\textsc{Par}$ rule could be proved by {\lemmaprefix} \ref{lm:semaitc_compositional} as the same way in \cite{Xu97}.

\section{Rely-Guarantee-based Reasoning of IFS}
\label{sect:rgifs}
This section presents the new forms of unwinding conditions and their soundness for IFS. Then, we show the compositionality of state-action based IFS for {\slang}.

We define the new forms of the locally respects and step consistent on events as follows. We assume a function $\Gamma: evts(\symbpes) \rightarrow RGCond$, where $RGCond$ is the type of the rely-guarantee specification, to specify the rely-guarantee specification of events in $\symbpes$. $G_{\Gamma(\symbevt)}$ is the guarantee condition in the rely-guarantee specification of the event $\symbevt$. Since the observation consistent condition has nothing to do with actions, we do not define a new form of this condition. 

\begin{definition}[Locally Respects on Events - LRE]
A parallel event system $\symbpes$ locally respects $\interf$ on events if 
\begin{equation*}
\begin{aligned}
\forall \symbevt \ \symbdomain \ \symbstate \ \symbstate' \ \symbcore. \ \symbevt \in evts(\symbpes) \wedge (\symbstate, \symbstate') \in G_{\Gamma(\symbevt)} \\
\longrightarrow (dom\_e(\symbstate, \symbcore, \symbevt) \ninterf d) \longrightarrow \symbstate \dsim{\symbdomain} \symbstate'
\end{aligned}
\end{equation*}
\end{definition}

\begin{definition}[Step Consistent on Events - SCE]
A parallel event system $\symbpes$ is step consistent on events if 
\begin{equation*}
\begin{aligned}
\forall \symbevt, \symbdomain, & \symbstate_1, \symbstate_2. \ \symbevt \in evts(\symbpes) \wedge \symbstate_1 \dsim{\symbdomain} \symbstate_2 \longrightarrow \\
&((dom\_e(\symbstate, \symbcore, \symbevt) \interf \symbdomain) \longrightarrow (\symbstate_1 \dsim{dom\_e(\symbstate, \symbcore, \symbevt)} \symbstate_2)) \longrightarrow \\
&
\begin{aligned}
(\forall \symbstate_1', \symbstate_2'. \ (\symbstate_1, \symbstate_1') \in G_{\Gamma(\symbevt)} \wedge (\symbstate_2, \symbstate_2') \in G_{\Gamma(\symbevt)} \\
\longrightarrow \symbstate_1' \dsim{\symbdomain} \symbstate_2')
\end{aligned}
\end{aligned}
\end{equation*}
\end{definition}

The locally respects condition requires that when an event $\symbevt$ executes, the modification of $\symbevt$ to the environment can affect only those domains which the domain executing $\symbevt$ is allowed to send information. The step consistent condition requires that the observation by a domain $\symbdomain$ when executing an event $\symbevt$ can depend only on $\symbdomain$'s observation before $\symbevt$ occurs, as well as the observation by the domain executing $\symbevt$ before $\symbevt$ occurs if that domain is allowed to send information to $\symbdomain$.  Different with the small-step UCs which examines each action in events in {\subsectprefix} \ref{subsec:small_ucs}, the event UCs consider the affect of events to the environment.

To prove the compositionality, we first show two lemmas as follows. {\lemmaprefix} \ref{lm:evtctx_consist} shows the consistency of the event context in computations of a closed $\symbpes$. {\lemmaprefix} \ref{lm:guar_consist} shows the compositionality of guarantee conditions of events in a valid and closed parallel event system. 

\begin{lemma}
\label{lm:evtctx_consist}
For any closed $\symbpes$, if events in $\symbpes$ are basic events, i.e., $\forall \symbevt \in evts(\symbpes). \ is\_basic(\symbevt)$,
then for any computation $\symbcomp$ of $\symbpes$, we have 

\begin{equation*}
\begin{aligned}
\forall i < len(\symbcomp) - 1, \symbcore. \ & (\exists \symbact. \ \symbcomp_i \trank{\symbact}{\symbcore} \symbcomp_{i+1}) \\
& \longrightarrow (\exists \symbevt \in evts(\symbpes). \ \symbevtctx_{\symbcomp_i}(\symbcore) = ev)
\end{aligned}
\end{equation*}

\end{lemma}

\begin{lemma}
\label{lm:guar_consist}
For any $\symbpes$, if 
\begin{itemize}
\item events in $\symbpes$ are basic events, i.e., $\forall \symbevt \in evts(\symbpes). \ is\_basic(\symbevt)$.
\item events in $\symbpes$ satisfy their rely-guarantee specification, i.e., $\forall \symbevt \in evts(\symbpes). \ \rgsat{\symbevt}{\Gamma(\symbevt)}$. 
\item $\rgsat{\symbpes}{\rgcond{\{\symbstate_0\}}{\varnothing}{UNIV}{UNIV}}$.
\end{itemize}
then for any computation $\symbcomp \in \compfun(\symbpes, \symbstate_0, \symbevtctx_0)$, we have 

\begin{equation*}
\begin{aligned}
\forall i < len(\symbcomp) - 1, \symbcore. \ & (\exists \symbact. \ \symbcomp_i \trank{\symbact}{\symbcore} \symbcomp_{i+1}) \\
& \longrightarrow (\symbstate_{\symbcomp_i}, \symbstate_{\symbcomp_{i+1}}) \in G_{\Gamma(\symbevtctx_{\symbcomp_i}(\symbcore))}
\end{aligned}
\end{equation*}

\end{lemma}

Based on the two lemmas, we have the following lemma for the soundness of event UCs, i.e., the conditions imply the small-step ones. 

\begin{lemma}[Soundness of Unwinding Conditions on Events]
For any $\symbpes$, if 
\begin{itemize}
\item $\symbconf_0 = (\symbpes, \symbstate_0, \symbevtctx_0)$.
\item events in $\symbpes$ are basic events, i.e., $\forall \symbevt \in evts(\symbpes). \ is\_basic(\symbevt)$.
\item events in $\symbpes$ satisfy their rely-guarantee specification, i.e., $\forall \symbevt \in evts(\symbpes). \ \rgsat{\symbevt}{\Gamma(\symbevt)}$. 
\item $\rgsat{\symbpes}{\rgcond{\{\symbstate_0\}}{\varnothing}{UNIV}{UNIV}}$.
\end{itemize}
then $\symbSM=\langle \symbConf, \symbAction, step, \symbconf_0 \rangle$, which is constructed according to {\defprefix} \ref{def:statemachine}, satisfies that 
\[
LRE \Longrightarrow LR \quad and \quad SCE \Longrightarrow SC
\]
\end{lemma}

We require that all events in $\symbpes$ are basic events to ensure the event context in computations of $\symbpes$ is consistent. It is reasonable since anonymous events are only used to represent the intermediate specification during execution of events. The last assumption is a highly relaxed condition and is easy to be proved. First, we only consider closed concurrent systems starting from the initial state $\symbstate_0$. Thus, the pre-condition only has the initial state and the rely condition is empty. Second, we concerns the environment affect of an event to other events, but not the overall modification, and thus the guarantee condition is the universal set. Third, IFS only concerns the action transition, but not the final state. Thus, the post-condition is the universal set.  


From this lemma and the small-step unwinding theorems ({\theoremprefix}s \ref{thm:ut_noninfl} and \ref{thm:ut_nonlk}), we have the compositionality of IFS as follows. 

\begin{theorem}[Compositionality of IFS]
\label{thm:comp_ifs}
For any $\symbpes$, if 
\begin{itemize}
\item $\symbconf_0 = (\symbpes, \symbstate_0, \symbevtctx_0)$.
\item events in $\symbpes$ are basic events, i.e., $\forall \symbevt \in evts(\symbpes). \ is\_basic(\symbevt)$.
\item events in $\symbpes$ satisfy their rely-guarantee specification, i.e., $\forall \symbevt \in evts(\symbpes). \ \rgsat{\symbevt}{\Gamma(\symbevt)}$. 
\item $\rgsat{\symbpes}{\rgcond{\{\symbstate_0\}}{\varnothing}{UNIV}{UNIV}}$.
\end{itemize}
then $\symbSM=\langle \symbConf, \symbAction, step, \symbconf_0 \rangle$, which is constructed according to {\defprefix} \ref{def:statemachine}, satisfies that 
\[
OC \wedge LRE \wedge SCE \Longrightarrow noninfluence
\]
and
\[
OC \wedge LRE \wedge SCE \Longrightarrow nonleakage
\]
\end{theorem}

By this theorem and {\lemmaprefix} \ref{lm:sm_equiv}, we provide a compositional approach of IFS for {\slang}. 



\section{Verifying IFS of Multicore Separation Kernels}
\label{sect:case}
By the proposed compositional approach for verifying IFS and its implementation in Isabelle/HOL, we develop a formal specification and its IFS proof of multicore separation kernels in accordance with the ARINC 653 standard. In this section, we use the concrete syntax created in Isabelle to represent the formal specification. 

\subsection{Architecture of Multicore Separation Kernels}

\begin{figure}[t]
\begin{center}
\includegraphics[width=3.2in]{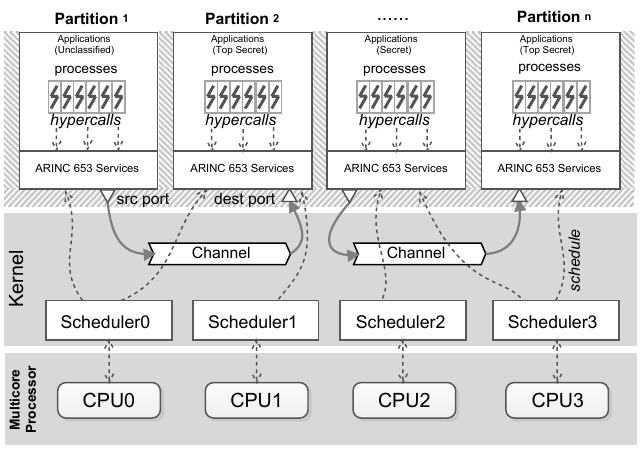}
\end{center}
\caption{Architecture of Multicore Separation Kernels}
\label{fig:arch}
\end{figure}

The ARINC 653 standard - Part 1 in Version 4 \cite{ARINC653p1_4} released in 2015 specifies the baseline operating environment for application software used within Integrated Modular Architecture on a multicore platform. It defines the \emph{system functionality} and requirements of \emph{system services} for separation kernels. 
As shown in {\figprefix} \ref{fig:arch}, separation kernels in multicore architectures virtualise the available CPUs offering to the partitions virtual CPUs. A partition can use one or more virtual CPUs to execute the internal code. 
Separation kernels schedule partitions in a fixed, cyclic manner. 

Information-flow security of separation kernels is to assure that there are no channels for information flows between partitions other than those explicitly provided. The security policy used by separation kernels is the \emph{Inter-Partition Flow Policy} (IPFP), which is intransitive. It is expressed abstractly in a partition flow matrix $\mathbf{partition\_flow}: partition \times partition \rightarrow mode$, whose entries indicate the mode of the flow. For instance, $\mathbf{partition\_flow}(P_1,P_2) = SAMPLING$ means that a partition $P_1$ is allowed to send information to a partition $P_2$ via a sampling-mode channel which supports multicast messages. 

\subsection{System Specification}
As a study case, the formal specification only considers the partitions, partition scheduling, and inter-partition communication (IPC) by sampling channels. We assume that the processor has two cores, $\symbcore_0$ and $\symbcore_1$.
A partition is basically the same as a program in a single application environment. Partitions have access to channels via \emph{ports} which are the endpoints of channels.
A significant characteristic of ARINC 653 is that the basic components are statically configured at built-time. The configuration is defined in Isabelle as follows. We create a constant $conf$ used in events. $c2s$ is the mapping from cores to schedulers and is bijective. $p2c$ is the deployment of partitions to schedulers and a partition could execute on some cores concurrently. A set of configuration constraints are defined to ensure the correctness of the system configuration. 
The kernel state defined as follows concerns states of schedulers and channels. The state of a scheduler shows which is the currently executing partition. The state of a channel is mainly about messages in its one-size buffer. 

\begin{isabellec}
\isacodeftsz
\ \ \ \ \isacommand{record}\isamarkupfalse%
\ Config {\isacharequal}\ c2s\ {\isacharcolon}{\isacharcolon}\ Core\ {\isasymRightarrow} \ Sched
\ \ \ \ p2s\ {\isacharcolon}{\isacharcolon}\ Part\ {\isasymRightarrow} \ Sched \ set \isanewline
\ \ \ \ \ \ \ \ \ \ \ \ \ \ \ \ \ \ \ \ \ \ \ \ \ \ \ \ \ \ \ \
p2p \ {\isacharcolon}{\isacharcolon}\ Port \ {\isasymRightarrow} \ Part \isanewline
\ \ \ \ \ \ \ \ \ \ \ \ \ \ \ \ \ \ \ \ \ \ \ \ \ \ \ \ \ \ \ \
scsrc \ {\isacharcolon}{\isacharcolon}\ SampChannel \ {\isasymRightarrow} \ Port \isanewline
\ \ \ \ \ \ \ \ \ \ \ \ \ \ \ \ \ \ \ \ \ \ \ \ \ \ \ \ \ \ \ \
scdests \ {\isacharcolon}{\isacharcolon}\ SampChannel \ {\isasymRightarrow} \ Port \ Set

\ \ \ \ \isacommand{axiomatization} conf :: Config \isanewline

\ \ \ \ \isacommand{record}\isamarkupfalse%
\ State {\isacharequal}\ cur \ {\isacharcolon}{\isacharcolon}\ Sched {\isasymRightarrow} \ Part \isanewline
\ \ \ \ \ \ \ \ \ \ \ \ \ \ \ \ \ \ \ \ \ \ \ \ \ \ \ \ \ \ 
schan \ {\isacharcolon}{\isacharcolon}\ SampChannel \ {\isasymrightharpoonup} \ Message \isanewline
\end{isabellec}

\begin{figure}[t]
\begin{isabellec}
\footnotesize

\isacommand{EVENT} Schedule \ ps @ $\symbcore$ \isacommand{WHERE} \isanewline
\ \ \ $ps \ typeof \ [] $ \isanewline
\isacommand{THEN} \isanewline
\ \ \ cur := cur \ ((c2s conf) $\symbcore$ := SOME p. (c2s conf) $\symbcore$ $\in$ (p2s conf) p ) \isanewline
\isacommand{END} \isanewline

\isacommand{EVENT} Write\_Sampling\_Message \ ps @ $\symbcore$ \isacommand{WHERE} \isanewline
\ \ \ $ps \ typeof \ [PORT,MSG] $ $\wedge$ is\_src\_sampport \ conf \ (ps!0) \isanewline
\ \ \ $\wedge$ (p2p \ conf) \ (ps!0) \ (cur \ (gsch \ conf \ $\symbcore$)) \isanewline
\isacommand{THEN} \isanewline
\ \ \ schan := schan (ch\_srcsampport \ conf (ps!0) := Some (ps!1))  \isanewline
\isacommand{END} \isanewline

\isacommand{EVENT} Read\_Sampling\_Message \ ps @ $\symbcore$ \isacommand{WHERE} \isanewline
\ \ \ $ps \ typeof \ [PORT] $ $\wedge$ is\_dest\_sampport \ conf \ (ps!0) \isanewline
\ \ \ $\wedge$ (p2p \ conf) \ (ps!0) \ (cur \ (gsch \ conf \ $\symbcore$)) \isanewline
\isacommand{THEN} \isanewline
\ \ \ SKIP  \isanewline
\isacommand{END} \isanewline

\isacommand{EVENT} Core\_Init \ ps @ $\symbcore$ \isacommand{WHERE} \isanewline
\ \ \ True \isanewline
\isacommand{THEN} \isanewline
\ \ \ SKIP \isanewline
\isacommand{END}

\end{isabellec}
\caption{Events Defined in the Specification}
\label{fig:evts_def}
\end{figure}

We define a set of events for separation kernels as shown in {\figprefix} \ref{fig:evts_def}. The $Schedule$ event chooses a partition deployed on core $\symbcore$ as the current partition of the scheduler of $\symbcore$. The $Write\_Sampling\_Message$ event updates the message buffer of a sampling, source port that belongs to the partition in which the event is executing. The $Read\_Sampling\_Message$ event does not change the kernel state. We specify a $SKIP$ action in the $Core\_Init$ event to initialize each core. However, we use this event to illustrate the usage of the \emph{event sequence}. 
The parallel event system in {\slang} of multicore separation kernels is thus defined as follows. The event systems defined on each core are the same. 

\begin{equation*}
\begin{aligned}
XKernel \equiv (\lambda \symbcore. \ Core\_Init; (Schedule \evtcomp \\
Write\_Sampling\_Message \evtcomp \\
Read\_Sampling\_Message))
\end{aligned}
\end{equation*}

We define a set of security domains for separation kernels. Each partition is a security domain. We define a security
domain for the scheduler on each core, which cannot be interfered by any other domains to ensure that the scheduler does not leak information via its scheduling decisions. The domain of events is dependent with the kernel state and defined as follows. The domain of $Write\_Sampling\_Message$ and $Read\_Sampling\_Message$ is the current partition on core $\symbcore$ on which the events are executing. The domain of $Schedule$ and $Core\_Init$ is the scheduler on core $\symbcore$ on which the events are executing. 

\begin{isabellec}
\footnotesize
\isacommand{definition}\isamarkupfalse%
\ dom\_of\_evt\ {\isacharcolon}{\isacharcolon}\ {\isachardoublequoteopen}State\ {\isasymRightarrow}\ Core \ {\isasymRightarrow}\ Event \ {\isasymRightarrow} Domain {\isachardoublequoteclose} \isanewline
\ \ \ \isakeyword{where}\ {\isachardoublequoteopen}dom\_of\_evt\ s \ k \ e {\isasymequiv}\isanewline
\ \ \ \ \ \ \textbf{if} e = Write\_Sampling\_Message \ \isasymor \isanewline
\ \ \ \ \ \ \ \ \ \  e = Read\_Sampling\_Message \ \textbf{then} \isanewline
\ \ \ \ \ \ \ \ \ \ \ \ \ \ \ \ \ \ \ \ \ \ \ \ \ \ ((cur \ s) \ ((c2s \ conf) \ k)) \isanewline
\ \ \ \ \ \ \textbf{else\ if}\ e = Schedule \isasymor \ e = Core\_Init \ \textbf{then}\ ((c2s \ conf) \ k) \isanewline
\ \ \ \ \ \ \textbf{else}\ ((c2s \ conf) \ k){\isachardoublequoteclose}
\end{isabellec}

The security policy is defined according to the channel configuration. If there is a channel from a partition $p_1$ to a partition $p_2$,
then $p_1 \interf p_2$. Since the scheduler can possibly schedule all partitions deployed on it, it can interfere with them. The interference relation of domains is defined as follows. 

\begin{isabellec}
\footnotesize
\isacommand{definition}\isamarkupfalse%
\ interf\ {\isacharcolon}{\isacharcolon}\ {\isachardoublequoteopen}Domain\ {\isasymRightarrow}\ Domain\ {\isasymRightarrow}\ bool{\isachardoublequoteclose}\ {\isacharparenleft}{\isachardoublequoteopen}{\isacharparenleft}{\isacharunderscore}\ {\isasymleadsto}\ {\isacharunderscore}{\isacharparenright}{\isachardoublequoteclose}{\isacharparenright}\isanewline
\ \ \ \isakeyword{where}\ {\isachardoublequoteopen}interf\ d1\ d2\ {\isasymequiv}\isanewline
\ \ \ \ \ \ \textbf{if}\ d1\ {\isacharequal}\ d2\ \textbf{then}\ True\isanewline
\ \ \ \ \ \ \textbf{else\ if}\ part\_on\_core \ conf \ d2 \ d1\ \textbf{then}\ True\ \isanewline
\ \ \ \ \ \ \textbf{else\ if}\ ch\_conn2 \ conf \ d1 \ d2 \ \textbf{then}\ True \isanewline
\ \ \ \ \ \ \textbf{else}\ False{\isachardoublequoteclose}
\end{isabellec}

The state observation of a domain to a state is defined as follows. The observation of a scheduler $d$ is that which is the currently executing partition on $d$. The observation of a partition $p$ is the message of channels with which the ports belonging to $p$ connect. Then, the equivalence relation $\dsim{\symbdomain}$ on states is defined as $state\_equiv$. 
\begin{isabellec}
\footnotesize

\isacommand{definition}\isamarkupfalse%
\ state\_obs\_sched\ {\isacharcolon}{\isacharcolon}\ {\isachardoublequoteopen}State \ {\isasymRightarrow}\ Part \ {\isasymRightarrow}\ State{\isachardoublequoteclose} \isanewline
\ \ \ \isakeyword{where}\ {\isachardoublequoteopen}state\_obs\_sched\ s\ d\ {\isasymequiv} 
s0{\isasymlparr} cur := (cur \ s0) \ (d:=(cur \ s) \ d) {\isasymrparr}
{\isachardoublequoteclose}\isanewline

\isacommand{definition}\isamarkupfalse%
\ state\_obs\_part\ {\isacharcolon}{\isacharcolon}\ {\isachardoublequoteopen}State \ {\isasymRightarrow}\ Part \ {\isasymRightarrow}\ State{\isachardoublequoteclose} \isanewline
\ \ \ \isakeyword{where}\ {\isachardoublequoteopen}state\_obs\_part\ s\ p\ {\isasymequiv} 
s0{\isasymlparr} schan := schan\_obs\_part \ s \ p {\isasymrparr}
{\isachardoublequoteclose}\isanewline

\isacommand{primrec}\isamarkupfalse%
\ state\_obs\ {\isacharcolon}{\isacharcolon}\ {\isachardoublequoteopen}State \ {\isasymRightarrow}\ Domain\ {\isasymRightarrow}\ State{\isachardoublequoteclose}\ \isanewline
\ \ \ \isakeyword{where}\ {\isachardoublequoteopen}state\_obs\ \ s \ (P \ p)  = state\_obs\_part \ s \ p {\isachardoublequoteclose} {\isacharbar} \isanewline
\ \ \ \ \ \ \ \ \ \ \ \ \ \ {\isachardoublequoteopen}state\_obs\ \ s \ (S \ p)  = state\_obs\_sched \ s \ p {\isachardoublequoteclose} \isanewline

\isacommand{definition}\isamarkupfalse%
\ state\_equiv \ {\isacharcolon}{\isacharcolon}\ {\isachardoublequoteopen}State \ {\isasymRightarrow}\ Domain\ {\isasymRightarrow}\ State \ {\isasymRightarrow} \ bool {\isachardoublequoteclose}\isanewline
\ \ \ \isakeyword{where}\ {\isachardoublequoteopen}state\_equiv\ s \ d \ t\ {\isasymequiv}\ state\_obs \ s \ d = state\_obs \ t \ d{\isachardoublequoteclose}
\end{isabellec}

\subsection{Rely-guarantee Specification of Events}
In order to compositionally reason IFS of the formal specification, we first present the rely-guarantee specification of events as illustrated in {\figprefix} \ref{fig:rgspecs}. The expressions $\llbrace \phi \rrbrace$ are concrete syntax for the set of states (or pairs of states) satisfying $\phi$. We present the value of a variable $x$ in $\phi$ after a transition by $x'$. 
 
The rely condition of $Schedule$ on $\symbcore_0$ shows that the event relies on that the currently executing partition on $\symbcore_0$ is not changed by the environment. The guarantee condition means that the execution of $Schedule$ does not change the currently executing partition on $\symbcore_1$. The execution $Write\_Sampling\_Message$ on $\symbcore_0$ also relies on that the currently executing partition on $\symbcore_0$ is not changed by the environment. The guarantee condition means that the execution of $Write\_Sampling\_Message$ does not change the currently executing partition on each core and other channels except the one connects the operated source port are not changed. 
The execution of $Read\_Sampling\_Message$ on $\symbcore_0$ also relies on the same condition, but does not change the kernel state. 

\begin{figure}
\footnotesize
\textbf{The Event $Schedule$ on $\symbcore_0$:}
\begin{equation*}
\begin{aligned}
\langle \llbrace True \rrbrace, \llbrace cur' \ (c2s \ conf \ \symbcore_0) = cur (c2s \ conf \ \symbcore_0) \rrbrace, \\
\llbrace cur' \ (c2s \ conf \ \symbcore_1) = cur (c2s \ conf \ \symbcore_1) \rrbrace, \llbrace True \rrbrace \rangle
\end{aligned}
\end{equation*}

\textbf{The Event $Write\_Sampling\_Message$ on $\symbcore_0$:}
\begin{equation*}
\begin{aligned}
\langle \llbrace True \rrbrace, \llbrace cur' \ (c2s \ conf \ \symbcore_0) = cur \ (c2s \ conf \ \symbcore_0) \rrbrace, \\
\llbrace cur' = cur \wedge (\exists ps. (\exists ! ch. \ ch = (ch\_srcsport \ conf \ (ps!0)) \\
\wedge (\forall ch1. \ (ch1 \neq ch \longrightarrow schan' \ ch1 = schan \ ch1))) \rrbrace, \\
\llbrace True \rrbrace \rangle
\end{aligned}
\end{equation*}

\textbf{The Event $Read\_Sampling\_Message$ on $\symbcore_0$:}
\begin{equation*}
\begin{aligned}
\langle \llbrace True \rrbrace, \llbrace cur' \ (c2s \ conf \ \symbcore_0) = cur \ (c2s \ conf \ \symbcore_0) \rrbrace, \\
\llbrace cur' = cur \wedge schan' = schan \rrbrace, \llbrace True \rrbrace \rangle
\end{aligned}
\end{equation*}


\caption{Rely-Guarantee Specification of Events in Separation Kernels}
\label{fig:rgspecs}
\end{figure}

\subsection{Security Proof}
According to {\lemmaprefix} \ref{thm:comp_ifs}, to show the information-flow security of our formal specification we only need to prove the assumptions of this theorem and that the events satisfy the event UCs. 
The first assumption of {\lemmaprefix} \ref{thm:comp_ifs} is satisfied on the state machine straightforwardly. The second one is trivial. The third and fourth ones are proved by the rely-guarantee proof rules defined in {\figprefix} \ref{fig:proofrule}. 

Next, we have to show satisfaction of event UCs in the formal specification. For each event in the formal specification, we prove that it satisfies the event UCs.

\section{Related Work and Conclusion}
\label{sect:rw_concl}
\paragraph{Rely-guarantee method.}
Initially, the rely-guarantee method for shared variable concurrent programs is to establish a post-condition for final states of terminated computations \cite{Xu97}. The languages used in rely-guarantee methods (e.g., \cite{Jones83,Xu97,Nieto03,LiangFF12}) are basically imperative programming languages with concurrent extensions (e.g., parallel composition, and $await$ statement). In this paper, we propose a rely-guarantee proof system for an event-based language, which incorporates the elements of system specification languages into existing rely-guarantee languages. We employee ``Events'' \cite{Abrial07} into rely-guarantee and provide event systems and parallel composition of them to model single-processing and concurrent systems respectively. Our proposed language enables rely-guarantee-based compositional reasoning at the system level.

Event-B \cite{Abrial07} is a refinement-based formal method for system-level modeling and analysis. In a machine in Event-B, the execution of an event, which describes a certain observable transition of the state variables, is considered to be atomic and takes no time. The parallel composition of Event-B models is based on shared events \cite{Silva12}, which can be considered as in message-passing manner. In \cite{Hoang10}, the authors extend Event-B to mimic rely-guarantee style reasoning for concurrent programs, but not provide a rely-guarantee framework for Event-B. In this paper, {\slang} is a language for shared variable concurrent systems. {\slang} provides a more expressive language than Event-B for the body of events. The execution of events in {\slang} is not necessarily atomic and we provide a rely-guarantee proof system for events. 


\paragraph{Formal verification of information-flow security.}
Formal verification of IFS has attracted many research efforts in recent years. Language-based IFS \cite{sabel03} defines security policies on programming languages and concerns the data confidentiality among program variables. The compositionality of language-based IFS has been studied (e.g. \cite{Mantel11,Murray16}). But for security at the system levels, the secrecy of actions is necessary such as for operating system kernels. 
State-action based IFS is formalized in \cite{rushby92} on a state machine and generalized and extended in \cite{Oheimb04} by nondeterminism. The IFS properties in \cite{rushby92,Oheimb04} are defined and verified on the seL4 microkernel. 
However, the compositionality of state-action based IFS \cite{rushby92,Oheimb04,Murray12} has not been studied in the literature. 

Recently, formal verification of microkernels and separation kernels is considered as a promising way for high-assurance systems \cite{Klein09b}. Information-flow security has been formally verified on the seL4 microkernel\cite{Murray13}, PROSPER hypervisor \cite{dam13}, ED separation kernel \cite{Heitmeyer08}, ARINC 653 standard \cite{Zhao16}, and INTEGRITY-178 \cite{richards10}, etc. 
In \cite{Murray12,Murray13,Zhao16}, the IFS properties are dependent with separation kernels, i.e., there is a specific security domain (\emph{scheduler}) in the definition of the properties. In our paper, the IFS properties are more general and we do not need to redefine new IFS properties in our study case. 
On the other hand, all these efforts are enforced on monocore kernels. Latest efforts on this topic aim at interruptable OS kernels, e.g., \cite{Chen16,Xu16}. However, formal verification of multicore kernels is still challenging. Although the formal specification is very abstract, we present the first effort of using the rely-guarantee method to compositional verification of multicore kernels in the literature. 

\paragraph{Discussion.}
Although, we only show the compositional reasoning of IFS by the rely-guarantee proof system in this paper, it is possible to use the proof system for the functional correctness and safety of concurrent systems. Invariants of concurrent systems could be compositionally verified by the rely-guarantee specification of events in the system. Deadlock-free of a concurrent system is possible to be verified by the pre- and post-conditions of events. For the functional correctness, we may extend the superposition refinement \cite{Back96} by considering the rely-guarantee specification to show that a concrete event preserves the refined one. This is one of our future work. 

By an implicit transition system in the semantics of an event system in {\slang}, events provide a concise way to define the system behavior. \citet{Abrial10} introduces a method to represent sequential programs by event-based languages. Based on this method and the concurrent statements in {\slang}, concurrent programs in other rely-guarantee methods can also be expressed by {\slang}. 

By the state machine representation of {\slang}, any state-action based IFS properties can be defined and verified in {\slang}. In this paper, we create a nondeterministic state machine from {\slang}, but we use the deterministic forms of IFS properties in \cite{Oheimb04} since the nondeterministic forms are not refinement-closed. This is also followed in \cite{Murray12} for seL4. 

discussion on the guard, pre-condition, and the guarantee condition.

\paragraph{Conclusion and future work. }
In this paper, we propose a rely-guarantee-based compositional reasoning approach for verifying information-flow security of concurrent systems. We design the {\slang} language, which incorporates the concept of ``Events'' into concurrent programming languages. We define the information-flow security and develop a rely-guarantee proof system for {\slang}. For the compositionality of IFS, we relax the atomicity constraint on the unwinding conditions and define new forms of them on the level of events. Then, we prove that the new unwinding conditions imply the security of {\slang}. The approach proposed in this paper has been mechanized in the Isabelle/HOL theorem prover. Finally, we create a formal specification for multicore separation kernels and prove the information-flow security of it. 
In the future, we would like to further study the refinement in {\slang} and the information-flow security preservation during the refinement. Then, we will create a complete formal specification for multicore separation kernels according to ARINC 653 and use the refinement to create a model at the design level.

\acks
We would like to thank Jean-Raymond Abrial and David Basin of ETH Zurich, Gerwin Klein and Ralf Huuck of NICTA, Australia for their suggestions.

%

\bibliographystyle{abbrvnat}
\bibliography{paperbibtex}

\end{document}